# Early- and late-time prediction of counter-current spontaneous imbibition, scaling analysis and estimation of the capillary diffusion coefficient

Pål Østebø Andersen [1,*]
1 Department of Energy Resources, University of Stavanger, 4021 Norway
* Correspondence: pal.andersen@uis.no
ORCID: 0000-0002-8552-094X

## Abstract

Solutions are investigated for 1D linear counter-current spontaneous imbibition (COUSI). The diffusion problem is scaled to depend only on a normalized coefficient $\Lambda_n(S_n)$ with mean 1 and no other parameters.

A dataset of 5500 functions $\Lambda_n$ was generated using combinations of (mixed-wet and strongly water-wet) relative permeabilities, capillary pressure and mobility ratios. Since the possible variation in $\Lambda_n$ appears limited (mean 1, positive, zero at $S_n = 0$, one maximum) the generated functions span most relevant cases. The scaled diffusion equation was solved for all 5500 cases and recovery profiles were analyzed in terms of time scales and early- and late time behavior.

Scaled recovery falls exactly on the square root curve $RF = T_n^{0.5}$ at early time. The scaled time $T_n = t/\tau T_{ch}$ accounts for system length $L$ and magnitude $\overline{D}$ of the unscaled diffusion coefficient via $\tau = L^2/\overline{D}$, and $T_{ch}$ accounts for $\Lambda_n$.

Scaled recovery was characterized by $RF_{tr}$ (highest recovery reached as $T_n^{0.5}$) and $lr$, a parameter controlling the decline in imbibition rate afterwards. This correlation described the 5500 recovery curves with mean $R^2 = 0.9989$. $RF_{tr}$ was 0.05 to 0.2 units higher than recovery when water reached the no-flow boundary.

The shape of $\Lambda_n$ was quantified by three fractions $z_{a,b}$. The parameters describing $\Lambda_n$ and recovery were correlated which permitted to (1) accurately predict full recovery profiles (without solving the diffusion equation); (2) predict diffusion coefficients explaining experimental recovery; (3) explain the combined impact of interactions between wettability / saturation functions, viscosities and other input on early- and late time recovery behavior.

## Article highlights

- Scaled recovery is characterized by level reached in root of time regime and the imbibition rate decline afterwards.
- The magnitude and shape of the diffusion coefficient and the system length is enough to predict recovery profiles.
- The capillary diffusion coefficient can be estimated accurately based on imbibition recovery data.

## Acknowledgments

The author acknowledges the Research Council of Norway and the industry partners of NCS2030 – RCN project number 331644 – for their support.

## Nomenclature

### Roman

| | | |
|---|---|---|
| $A$ | = | Scaled flux proportionality coefficient, - |
| $D$ | = | Capillary diffusion coefficient, m²/s |
| $f_w$ | = | Fractional water flow function, - |
| $J$ | = | Scaled capillary pressure, - |
| $K$ | = | Permeability, m² |
| $L$ | = | System length, m |
| $m$ | = | Saturation profile exponent, - |
| $n_i$ | = | Corey exponent, - |
| $p_i$ | = | Phase pressure, Pa |
| $P_c$ | = | Capillary pressure, Pa |
| $r, lr$ | = | Late time recovery tuning parameter (and its logarithm), - |
| $RF$ | = | Recovery factor, - |
| $RF_{cr}$ | = | Recovery when water reaches $X = 1$, - |
| $RF_{tr}$ | = | Recovery when linear trend with square root of time ends, - |
| $s_w$ | = | Water saturation |
| $S$ | = | Water saturation normalized over mobile saturation range, - |
| $S_n$ | = | Water saturation normalized over positive capillary pressure range, - |
| $t$ | = | Time, s |
| $T$ | = | Normalized time, - |
| $T_{ch}$ | = | Characteristic dimensionless time to collect curves to one line, - |
| $T_n$ | = | Normalized time, - |
| $u$ | = | Darcy velocity, m/s |
| $x$ | = | Length from open side, m |
| $X$ | = | Scaled length, - |
| $z_{a,b}$ | = | Fraction area of diffusion coefficient on upper half of the interval $a < S_n < b$, - |

### Greek

| | | |
|---|---|---|
| $\lambda_i$ | = | Phase mobility, 1 / (Pa s) |
| $\Lambda$ | = | Dimensionless capillary diffusion coefficient, - |
| $\Lambda_n$ | = | Normalized capillary diffusion coefficient with mean 1, - |
| $\tau$ | = | Time scale of recovery, s |
| $\mu_i$ | = | Phase viscosity, Pa s |
| $\sigma_{ow}$ | = | Interfacial tension, N / m |
| $\phi$ | = | Porosity, - |

## 1. Introduction

Spontaneous imbibition is a process where capillary forces cause uptake of wetting fluid (referred to as water) into a porous medium and simultaneous displacement of less or non-wetting fluid (referred to as oil). In the subsurface it is especially relevant for oil and gas recovery in naturally fractured reservoirs (Mason and Morrow 2013) and water uptake during hydraulic fracturing in shale (Makhanov et al. 2014; Li et al. 2019). In the former, injected water displaces oil from disconnected matrix blocks by spontaneous imbibition and gravity (Xie and Morrow 2001; Karimaie et al. 2006), while advective flow occurs in the permeable fracture network.

There are two main modes of spontaneous imbibition. The first is counter-current spontaneous imbibition (COUSI) which occurs in symmetrical systems where water surrounds the matrix block on all open sides: e.g. 1D systems with one side open or imbibition in cylindrical core plug samples, as used in the traditional Amott test (Amott 1959). COUSI is the focus of this work. The second mode is co-current spontaneous imbibition which occurs where parts of the open surface are exposed to water and the rest to oil (Hamon and Vidal 1986; Bourbiaux and Kalaydjian 1990). Then both phases flow mainly co-currently towards the oil exposed surfaces (Pooladi-Darvish and Firoozabadi 2000; Andersen 2021b). The oil production is predominantly co-current, but less favorable mobility ratio or higher oil-wetness can reduce this dominance (Andersen and Ahmed 2021).

Spontaneous imbibition is a strong indicator of wettability in the sense that the water uptake, and hence oil production, is limited by the degree of water-wetness (Kovscek et al. 1993; Zhou et al. 2000). If the rock is strongly oil-wet there is no uptake, while stronger water-wetness means more uptake (Anderson 1987a). A strongly water-wet (SWW) rock will produce as much oil by COUSI as forced imbibition. Capillarity is the driving force, which vanishes when zero capillary pressure has been reached. Capillary diffusion can also impact estimation of relative permeabilities and residual saturations in core flooding experiments (Rapoport and Leas 1953; Andersen 2021a, 2022) by liquid holdup from capillary forces.

Significant efforts have been made to understand how various properties affect COUSI and to upscale lab experiments. This has been done particularly considering the nonlinear partial differential equation describing COUSI. Permeability, porosity, block dimensions and shape and fluid properties such as viscosities and interfacial tension have been studied and built into time scales, usually considering a fixed rock type and wettability (Mattax and Kyte 1962; Hamon and Vidal 1986; Ma et al. 1997; Standnes 2006; Mason et al. 2009; Standnes and Andersen 2017). Relative permeability and capillary pressure functions also play an important role coupled with the viscosities. Zhou et al. (2002) included characteristic mobility end points in the time scale. Other works have included the role of viscosity or viscosity ratio using correction factors relative to established time scales (Fischer et al. 2008; Standnes 2009; Mason et al. 2010; Meng et al. 2017). The focus has also here usually been for SWW systems, scaling experiments and limited theoretical cases rather than solving the mathematical problem in general. At high Bond numbers, fluid densities and block height can also matter (Schechter et al. 1994; Xie and Morrow 2001; Bourbiaux 2009). Bourbiaux and Kalaydjian (1990) found that fluid mobilities during COUSI were necessarily lower than during co-current flow to explain experimental observations. Qiao et al. (2018) simulated their observations consistently by accounting for viscous coupling. Gravity and viscous coupling are not considered within the scope of this work.

Aronofsky et al. (1958) suggested that experimental imbibition recovery could be modeled as exponential with time. COUSI is usually not 1D and linear (in lab or field), but approximated so by introducing a characteristic length (Ma et al. 1997). This can affect the functional relation between recovery and time although Mason et al. (2009) indicated that the deviation from square root of time trends was not very significant. General solutions for 1D linear COUSI by McWhorter and Sunada (1990) accounted for arbitrary saturation functions and proved that recovery should follow a square root of time profile, but only until a no-flow boundary was reached. Their solution was later used by Schmid and Geiger (2013) to scale experimental imbibition data for all wetting conditions. It was extended by Andersen et al. (2020) to account for viscous coupling. It was also demonstrated that late time recovery does not scale to one curve. Ruth et al. (2007) estimated a self-similar variable as function of saturation and predicted square root of time recovery at early time. Before continuing, it should be stated that in tight media imbibition oil recovery can be proportional to time with exponents significantly different than 0.5 (Li et al. 2019). Our work focuses on conventional media where the standard capillary diffusion equation is considered representative.

Imbibition recovery at late time has been a challenge to predict without numerical simulation. Tavassoli et al. (2005) assumed Corey relative permeabilities and that the saturation profile was a polynomial function of distance with time dependent coefficients. They derived recovery solutions at early and late time, the former following the square root of time. However, they did not get continuous slope in recovery at the transition time. Their solution was also independent of capillary pressure except for its derivative at the inlet saturation (we will show that to be incorrect). March et al. (2016) approximated the late time behavior with an exponential profile and optimized the transition by extending the early square root of time period (which is considered a good approximation) and ensuring equal recovery and slope at the transition to the exponential behavior. However, the exponential model did not represent late time recovery well in many of the cases. Andersen (2021c) studied shale gas recovery with mathematically same type equation as COUSI. It was shown that the linear trend of recovery with square root time could last significantly longer than the critical time; 0.2 to 0.4 units higher recovery was reached than at the critical time (where 1.0 equals full recovery), before the numerical solution differed by 2% from the straight line. Velasco-Lozano and Balhoff (2021) modified the method by Tavassoli et al. (2005). They assumed the saturation profile at the end of the infinite-acting period had same shape as an infinite-acting profile, but that all the imbibed water was within the core. By optimizing four tuning parameters in a polynomial saturation profile to fit the true profile at early time and demanding continuous derivative in recovery at the transition to finite-acting regime they could predict late time saturation profiles and recovery with time. Momeni et al. (2022) derived asymptotic late-time solutions for SWW cases.

From the above review we observe a few knowledge gaps: (a) Little has been said about possible distinctions between SWW and mixed-wet (MW) media regarding early and late time behavior; (b) Existing solutions for all-time recovery are either numerical; use non-general correlations (e.g. derived under limiting assumptions) with poor and unjustified transition to late time; and lack clear descriptions of what controls the late time behavior and its variations (c) The only truly general solution is semi-analytical in integral form, thus lacking intuitive explanatory power and is valid only until the no-flow boundary is met; (d) Existing late time correlations suffer from poor transition from early to late time or ability to follow the numerical solutions.

In this work we investigate recovery during COUSI accounting for 'all' conditions (parameters related to wettability / saturation functions, core- and fluid parameters, but not heterogeneity or dynamic changes in the stated parameters). It is shown that only a normalized capillary diffusion coefficient (CDC) function $\Lambda_n$ with mean 1 is needed to determine scaled solutions (**Section 2**), while specific case solutions follow from unscaling. Recovery behavior is investigated based on solving the diffusion equation with 5500 $\Lambda_n$ different functions, expected to cover most possible scenarios. A correlation for early and late time recovery containing two tuning parameters (**Section 3**) is found to represent scaled recovery very well for the simulated cases and is used to describe a given recovery profile. Three parameters that describe how $\Lambda_n$ is shaped are quantified (**Section 4**) and correlated with the two recovery profile parameters. That allows us to estimate recovery profiles directly from a given CDC, and the other way around; estimate the CDC from recovery data, which is illustrated with examples on literature data (**Section 5**). In addition to the stated knowledge gaps, some questions we address are: How does the CDC affect recovery behavior; How does wettability and viscosity ratio affect imbibition recovery; How can we use experimental recovery data from COUSI to estimate the CDC; What determines when recovery no longer acts proportional to the square root of time?

## 2. Mathematical description of COUSI

### 2.1. Geometry and mass balance equations

We consider 1D linear, incompressible, immiscible flow of oil ($o$) and water ($w$), **Figure 1**, expressed with phase saturations $s_i$ and phase pressures $p_i$ ($i = o, w$). The variables are constrained by volume conservation and the imbibition capillary pressure function $P_c(s_w)$:

(1) $$s_w + s_o = 1, \qquad p_o - p_w = P_c(s_w)$$

Fluid fluxes $u_i$ are described by Darcy's law. Gravity is ignored, and the system has constant porosity $\phi$, absolute permeability $K$ and wettability (represented by saturation functions). The system is open to water at $x = 0$ and closed at $x = L$. The open side has zero capillary pressure $P_c$ corresponding to a fixed saturation $s_w^{eq}$, defined such that $P_c(s_w^{eq}) = 0$ (Hamon and Vidal 1986; Bourbiaux and Kalaydjian 1990). The system is saturated with oil and immobile water, yielding a positive capillary pressure driving COUSI.

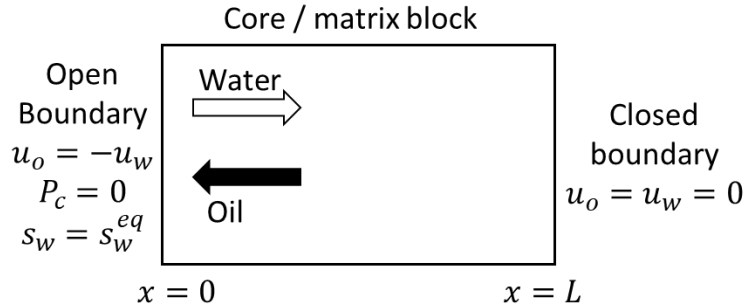

**Figure 1 Illustration of the system and the boundary conditions. Water and oil flow counter-currently through the open boundary at $x = 0$ by spontaneous imbibition.**

Under the stated assumptions, the COUSI system is described by the well-known nonlinear capillary diffusion equation (McWhorter and Sunada 1990; Tavassoli et al. 2005):

(2) $$\phi \frac{\partial s_w}{\partial t} = -K \frac{\partial}{\partial x}\left(\lambda_o f_w \frac{\partial P_c}{\partial x}\right), \qquad (0 < x < L).$$

$f_w$ is the fractional flow function, related to fluid mobilities $\lambda_i$ (relative permeability $k_{ri}$ divided by viscosity $\mu_i$) by:

(3) $$f_w = \frac{\lambda_w}{\lambda_w + \lambda_o}, \qquad \lambda_i = \frac{k_{ri}}{\mu_i}, \qquad (i = o, w).$$

The initial condition is uniform residual water saturation $s_{wr}$. This and the boundary conditions are expressed as:

(4) $$s_w(x, t = 0) = s_{wr}, \qquad s_w(x = 0, t) = s_w^{eq}, \qquad u_i(x = L, t) = 0$$

Capillary diffusion causes the saturations to approach $s_w^{eq}$ throughout the system after infinite time.

### 2.2. Scaled representation

We introduce scaling of saturation, spatial axis and time:

(5) $$S = \frac{s_w - s_{wr}}{\Delta s_w}, \qquad \Delta s_w = 1 - s_{or} - s_{wr}, \qquad X = \frac{x}{L}, \qquad T = \frac{t}{\tau}$$

$s_{or}$ is residual oil saturation, $\Delta s_w$ the range of mobile saturations and $\tau$ a time scale to be defined soon. The imbibition saturation functions are monotonous functions of $S$. Capillary pressure is expressed using the dimensionless $J$-function (Dullien 1992):

$$P_c = \sigma_{ow}\sqrt{\phi/K}\,J(S), \tag{6}$$

In line with Young-Laplace' equation, it states that capillary pressure increases with interfacial tension $\sigma_{ow}$ and inverse (characteristic) pore radius. From (2) we obtain a scaled transport equation:

$$\frac{\partial S}{\partial T} = \frac{\tau}{L^2}\frac{\partial}{\partial X}\left(D(S)\frac{\partial S}{\partial X}\right), \tag{7}$$

where $D(S)$ is a CDC with one part containing known constant parameters (unit m²/s), and a dimensionless saturation dependent part $\Lambda(S)$.

$$D(S) = \frac{\sigma_{ow}\sqrt{K/\phi}}{\mu_m \Delta s_w}\Lambda(S), \qquad \mu_m = (\mu_o \mu_w)^{0.5} \tag{8}$$

$$\Lambda(S) = \mu_m \lambda_o f_w\left(-\frac{dJ}{dS}\right) = \frac{k_{rw}k_{ro}\left(-\frac{dJ}{dS}\right)}{(\mu_o/\mu_w)^{0.5}k_{rw} + (\mu_w/\mu_o)^{0.5}k_{ro}} \tag{9}$$

$\Lambda(S)$ contains the saturation function terms multiplied by the mean viscosity. The latter makes $\Lambda(S)$ dimensionless and independent of $\mu_m$ (but dependent on the viscosity ratio $\mu_w/\mu_o$). The scaled initial and boundary conditions become:

$$S(X=0,T) = S_{eq} = \frac{s_w^{eq} - s_{wr}}{\Delta s_w}, \qquad \frac{\partial S}{\partial X}\Big|_{X=1} = 0, \qquad S(X,T=0) = 0 \tag{10}$$

Although the saturation functions span $0 < S < 1$, the relevant range for spontaneous imbibition is $0 < S < S_{eq}$. $S_{eq}$ can be less than 1 for cases that are not SWW and it is convenient to rescale the saturation:

$$S_n = \frac{S}{S_{eq}} = \frac{s_w - s_{wr}}{s_w^{eq} - s_{wr}} \tag{11}$$

Only saturations with positive capillary pressure, $0 < S_n < 1$, affect the solution. We update the transport equation (7):

$$\frac{\partial S_n}{\partial T} = \frac{\tau}{L^2}\frac{\partial}{\partial X}\left(D(S_n)\frac{\partial S_n}{\partial X}\right), \tag{12}$$

Now we select the time scale $\tau$ to account for the length $L$ and magnitude of $D$:

$$\tau = L^2/\bar{D} = \frac{\mu_m \Delta s_w L^2}{\sigma_{ow}\sqrt{K/\phi}\,\bar{\Lambda}}, \tag{13}$$

$$\bar{D} = \int_{S_n=0}^{1} D(S_n)dS_n, \qquad \bar{\Lambda} = \int_{S_n=0}^{1}\Lambda(S_n)dS_n \tag{14}$$

$\bar{D}$ and $\bar{\Lambda}$ denote averaged dimensional and dimensionless CDCs, respectively, over the saturation range with positive capillary pressure. Dividing either of the two CDCs by their mean gives the same normalized function $\Lambda_n(S_n)$ with mean 1.

$$\Lambda_n(S_n) = \frac{D(S_n)}{\bar{D}} = \frac{\Lambda(S_n)}{\bar{\Lambda}}. \tag{15}$$

With the time scale (13) we obtain the final scaled system of equations:

$$\frac{\partial S_n}{\partial T} = \frac{\partial}{\partial X}\left(\Lambda_n(S_n)\frac{\partial S_n}{\partial X}\right), \tag{16}$$

$$S_n(X=0,T) = 1, \qquad \frac{\partial S_n}{\partial X}\Big|_{X=1} = 0, \qquad S_n(X,T=0) = 0 \tag{17}$$

In this system we have normalized the independent variables space $0 < X < 1$ and time $T > 0$ and the dependent saturation variable $0 < S_n < 1$. The boundary conditions are the same for all cases. The only parameter affecting the scaled solution is the saturation dependent and normalized CDC $\Lambda_n(S_n)$ which is positive, has mean 1, is zero at $S_n = 0$ (and $S_n = 1$ for SWW cases) and is independent of mean viscosity. This represents *all 1D linear counter-current imbibition problems, for any input and wetting states*.

Oil recovery factor $RF$ (the fraction of oil that can be produced by COUSI) is given by:

| (18) | $$RF(T) = \bar{S}_n(T), \qquad \bar{S}_n = \int_{X=0}^{1} S_n(X,T) dX, \qquad 0 \leq RF(T) \leq 1.$$ |
|---|---|

$RF$ starts at 0 and ends at 1 (regardless of wettability and residual and initial oil saturation).

## 3. Theory motivated characterization of COUSI

### 3.1. Semi-analytical solution for early time

McWhorter and Sunada (1990) developed a semi-analytical solution to 1D COUSI under the limitation that the no-flow boundary had not been reached by imbibing water. Their solution was adapted to our normalized equations (16) and (17). For a full derivation, see **Supp Mat Section A**. Mainly, the position $X$ of saturation $S_n$, and recovery factor, are both proportional to $T^{0.5}$:

| (19) | $$X(S_n, T) = 2AF'(S_n)T^{0.5}, \qquad RF = 2AT^{0.5}$$ |
|---|---|

where $F'(S_n) = dF/dS_n$. The function $F(S_n)$ and the parameter $A$ can be calculated when $\Lambda_n$ is provided but they are defined implicitly in integral form and require numerical evaluation:

| (20) | $$F(S_n) = 1 - \left[\int_{\beta=S_n}^{1} (\beta - S_n) \frac{\Lambda_n(\beta)}{F(\beta)} d\beta\right] \cdot \left[\int_{\beta=0}^{1} \beta \frac{\Lambda_n(\beta)}{F(\beta)} d\beta\right]^{-1}$$ |
|---|---|
| (21) | $$A^2 = \frac{1}{2} \int_{\beta=0}^{1} \beta \frac{\Lambda_n(\beta)}{F(\beta)} d\beta$$ |

$\beta$ is an integration variable. The solution (19) to (21) is valid until the critical time $T_{cr}$ when the fastest saturation $S_n = 0$ meets the no-flow boundary $X(S_n = 0, T_{cr}) = 1$:

| (22) | $$T_{cr} = \frac{1}{4A^2} \frac{1}{[F'(S_n = 0)]^2}$$ |
|---|---|

We next scale $T$ with a factor $T_{ch}$, to yield same recovery when plotted against the resulting normalized time $T_n$:

| (23) | $$RF = T_n^{0.5}, \qquad T_n = \frac{T}{T_{ch}} = \frac{t}{\tau T_{ch}}, \qquad T_{ch} = \frac{1}{4A^2}$$ |
|---|---|

Especially, $RF$ equals $T_n^{0.5}$ at early time for all cases. From (22) and (23), recovery at the critical time is:

| (24) | $$RF_{cr} = T_{n,cr}^{0.5} = \sqrt{\frac{T_{cr}}{T_{ch}}} = \frac{1}{F'(S_n = 0)}$$ |
|---|---|

### 3.2. Transition from early to late time

Early time recovery $RF = T_n^{0.5}$ is linear plotted against $T_n^{0.5}$ and obeys $\frac{dRF}{d\sqrt{T_n}} = 1$ until $T_n = T_{n,cr}$. Considering recovery trends after critical time, this linear behavior appears to last significantly longer (March et al. 2016; Andersen 2021c). To capture this, we define a transition time $T_{n,tr} > T_{n,cr}$ when the slope deviates 'enough' from 1, here selected as when the slope has decreased to 0.9:

| (25) | $$\frac{dRF}{d\sqrt{T_n}}\bigg|_{T_n=T_{n,tr}} = 0.9,$$ |

The value 0.9 represented well where the numerically calculated recovery curves appeared to intersect the extended straight line $(RF = \sqrt{T_n})$, illustrated in **Section 5.3**. and improved the overall match of the investigated dataset compared to values such as 0.95 and 1 (no transition).

Since recovery until this point is well approximated by the straight line, we formulate it mathematically as:

| (26) | $$RF = T_n^{0.5}, \qquad (0 < T_n < T_{n,tr})$$ |

and we define recovery at the transition point $RF_{tr}$ as:

| (27) | $$RF_{tr} = T_{n,tr}^{0.5}$$ |

After this time, recovery deviates from the linear trend. As our main focus is on recovery we from now define late time as after $T_{n,tr}$, but remind that this generally occurs later than the critical time.

### 3.3. Late time recovery

The correlation derived by Tavassoli et al. (2005) for late time recovery will be assumed except in general form below with arbitrary constants $c_0, c_1, c_2, r$. We determine the constants to our own constraints.

| (28) | $$RF(T_n) = c_0 - \frac{c_1}{(T_n + c_2)^r}$$ |

Upon time differentiation, the expression is equivalent to Arp's harmonic decline curve (Arp 1945). We require that $RF$ equals 1 at infinite time; $RF$ is on the square root profile at $T_{n,tr}$; and the derivative of recovery wrt. $\sqrt{T_n}$ is 0.9 at $T_{n,tr}$:

| (29) | $$RF(T_n \to \infty) = 1, \qquad RF(T_{n,tr}) = RF_{tr}, \qquad \frac{dRF}{d\sqrt{T_n}}\bigg|_{T_n=T_{n,tr}} = 0.9$$ |

These constraints applied to (28) eliminates $c_0, c_1, c_2$, yielding the following late time expression:

| (30) | $$RF(T_n) = 1 - \left[1 + \frac{0.9}{2r}\frac{T_n - [RF_{tr}]^2}{RF_{tr} - [RF_{tr}]^2}\right]^{-r}(1 - RF_{tr}), \qquad (T_{n,tr} < T_n < \infty)$$ |

Given $RF_{tr}$; $r$ is the only tuning parameter for recovery between $T_{n,tr}$ and $T_n \to \infty$. For convenience, we will refer to the logarithmic value of $r$, called $lr \coloneqq \log_{10} r$. Sensitivity analysis showed that the correlation (30) converges to an exponential profile for $lr > 1.5$ (regardless value of $RF_{tr}$):

| (31) | $$RF = 1 + (RF_{tr} - 1)\frac{\exp(CT_n)}{\exp(CT_{n,tr})}, \qquad C = \frac{0.9}{2(RF_{tr} - 1)RF_{tr}}$$ |

### 3.4. Summarized parameterization of recovery correlation

Based on the previous discussion one parameter $RF_{tr}$ is used to describe early time recovery $RF(T_n)$ with (26), and a second parameter $lr$ is used to describe the continued late time profile using (30). The combined correlation is illustrated in **Figure 2** for a low and high $RF_{tr}$ and three choices of $lr$ each. At higher $lr$, the late time recovery more quickly reaches 1. The case with $lr = 1.5$ is indistinguishable from an exponential correlation. To get recovery against $T$ we need the factor $T_{ch}$. Recovery against regular time $t$ also requires $\tau$, see (23).

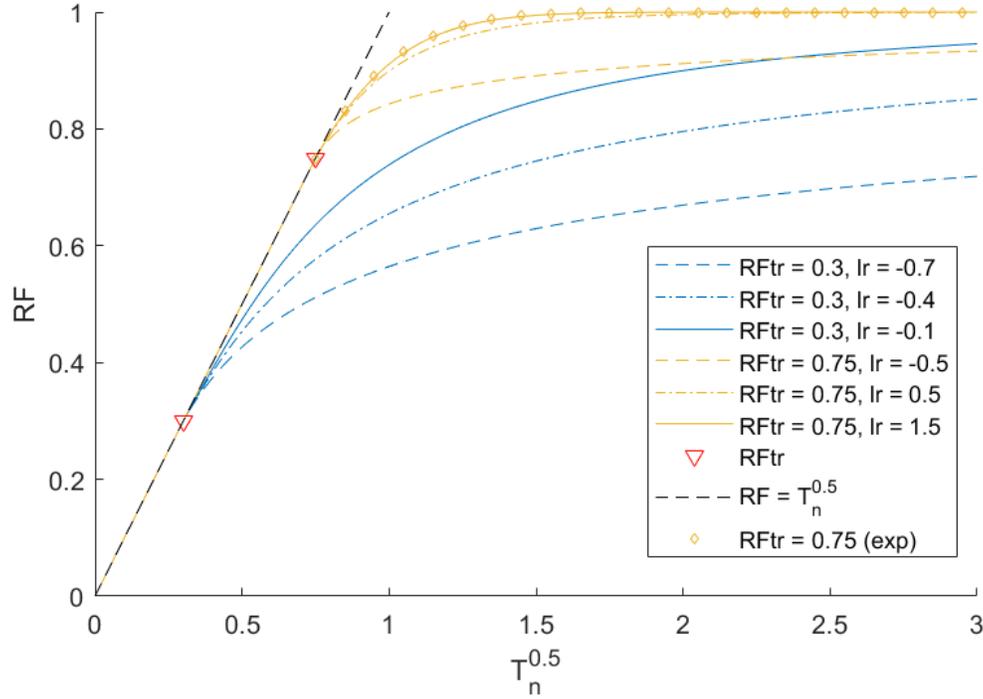

**Figure 2 Plot of the recovery correlation against $T_n^{0.5}$ for various $RF_{tr}$ and $lr$. At $lr > 1.5$ the curve converges to an exponential function (yellow points).**

### 3.5. Estimate of spatial saturation profiles before critical time

Before critical time, saturation positions are proportional to $T_n^{0.5}$ and a function $F'(S_n)$, see (19). $S_n = 1$ has zero speed, while $S_n = 0$ travels the fastest and reaches $X = 1$ at critical time, see (22). The following expression suggests a function $F'(S_n)$ (the factor to $T_n^{0.5}$) with an exponent $m$ determining the shape:

| (32) | $$X = \frac{1}{2}\left[(1 - S_n^m) + (1 - S_n)^{\frac{1}{m}}\right]\frac{T_n^{0.5}}{RF_{cr}}, \qquad (0 < T_n < T_{n,cr})$$ |
|---|---|

The averaged saturation profile must equal recovery, see (18). In particular, at critical time when $X(0) = 1$, we obtain $RF = RF_{cr}$. That allows to determine $m$ from the parameter $RF_{cr}$:

| (33) | $$m = \frac{RF_{cr}}{1 - RF_{cr}}$$ |
|---|---|

Thus, from estimates of $RF$ and $RF_{cr}$ we can estimate a saturation profile with correct amount imbibed water and front position.

## 4. Characterizing the scaled diffusion coefficient $\Lambda_n$

By reducing the COUSI problem to (16) and (17) through scaling, *the function $\Lambda_n$ is the only input affecting the solution*. For a simple description of a given function $\Lambda_n(S_n)$ we introduce a few parameters quantifying its shape. Define $z_{a,b}$ as the fraction area of $\Lambda_n$ on the interval $a < S_n < b$ which is on the upper half of that interval:

| (34) | $$z_{a,b} = \frac{\int_{S_n=\frac{a+b}{2}}^{b}\Lambda_n(S_n)dS_n}{\int_{S_n=a}^{b}\Lambda_n(S_n)dS_n}, \qquad (0 < z_{a,b} < 1)$$ |
|---|---|

A higher $z_{a,b}$ means $\Lambda_n$ is shifted more towards $b$ than $a$ on the interval. As the fraction only indicates shape, we could replace $\Lambda_n$ with $D$ in (34) and get the same answer. Three fractions $z_{0,1}, z_{0,0.5}, z_{0.5,1}$ are used to characterize each function $\Lambda_n$ in this work. Andersen (2021c) showed that $z_{0,1}$ correlated strongly with the solution output for a shale gas problem.

## 5. Results and discussion

### 5.1. Workflow and implementation

We select correlations representing relative permeability- and $J$-functions, which combined with e.g. mobility ratio in (9) and (15) yield realistic functions $\Lambda_n$ (**Section 5.2**). For a given $\Lambda_n(S_n)$ we calculate recovery $RF(T)$ numerically by solving the PDE (16), but also semi-analytically until critical time. The recovery behavior is parameterized with $RF_{cr}$ and $A$ from (20), (21) and (24) in the semi-analytical solution; and $RF_{tr}$ (defined by where $\frac{dRF}{d\sqrt{T_n}}=0.9$) and $lr$ by fitting (30) to $RF(T_n) > RF_{tr}$ from the numerical solution. The input $\Lambda_n$ is described with the three parameters $z_{0,1}, z_{0,0.5}, z_{0.5,1}$.

The numerical solution was implemented fully implicit in MATLAB as described in **Supp Mat Section B**. We used 500 equal grid cells and 50 000 time steps (equal on a square root axis) until $\sqrt{T_n} = 5$ (2 orders of magnitude higher than when $RF = 0.5$, and 3.4 orders higher than when $RF = 0.1$ in the linear regime). See convergence analyses in **Supp Mat Section C**.

The ability of the correlation (26) and (30) to describe recovery is demonstrated and the link between $\Lambda_n$ and recovery behavior is discussed in terms of the characterizing parameters for an example based on literature data (**Section 5.3**). The analysis is generalized by forming a dataset of 5500 functions $\Lambda_n$ and corresponding recovery curves where the characteristic parameters of each case are collected (**Section 5.4**). Relations between $\Lambda_n$ parameters and recovery parameters are demonstrated and quantified (**Section 5.5**). The results explain how functions $\Lambda_n$ translate into different early- and late time recovery. This is exemplified by predicting recovery without needing to solve the PDE for two literature data cases (**Section 5.6**); and to estimate CDCs explaining experimental recovery data (**Section 5.7**). The approaches are validated by comparing estimated recovery with the numerical solution.

### 5.2. Saturation function correlations and normalized CDCs

We use extended Corey function correlations (Brooks and Corey 1966) for relative permeability:

| (35) | $k_{ri} = k_{ri}^* S_i^{n_i}, \qquad n_i = n_{i1} S + n_{i2}(1 - S), \qquad (i = o, w)$ |

The exponents $n_i$ vary linearly with $S$ from $n_{i1}$ to $n_{i2}$ for flexibility. The $J$-function is a modified Bentsen and Anli (1976) correlation, where we have incorporated $J(S_{eq}) = 0$:

| (36) | $J(S) = -J_1 \ln\left(\dfrac{S}{S_{eq}}\right) + J_2 \ln\left(\dfrac{1-S}{1-S_{eq}}\right)$ |

The resulting CDC $\Lambda(S)$, defined by (9), becomes:

| (37) | $\Lambda(S) = \dfrac{J_2 k_{ro}^*}{\left(\dfrac{\mu_o}{\mu_w}\right)^{0.5}} \dfrac{S^{n_w-1}(1-S)^{n_o}\left(\dfrac{J_1}{J_2}\right) + S^{n_w}(1-S)^{n_o-1}}{S^{n_w} + \left(\dfrac{k_{ro}^*}{k_{rw}^*}\dfrac{\mu_w}{\mu_o}\right)(1-S)^{n_o}}$ |

At the end saturations $S = 0$ and $S = 1$, $\frac{dJ}{dS}$ goes to negative infinity, but $\Lambda(S)$ goes to 0 for typical Corey exponent values $n_i(S_i = 0) > 1$. The parameters $J_1, J_2, k_{rw}^*, k_{ro}^*, \mu_w, \mu_o$ affect the function shape through the two parameter ratios, $\frac{J_1}{J_2}$ and $\frac{k_{ro}^*}{k_{rw}^*}\frac{\mu_w}{\mu_o}$, while they only affect the magnitude if the ratios are kept constant.

The factor $\frac{J_2 k_{ro}^*}{\left(\frac{\mu_o}{\mu_w}\right)^{0.5}}$ cancels during normalization to $\Lambda_n(S_n) = \Lambda(S_n)/\int_{S_n=0}^{1} \Lambda(S_n)dS_n$ and does not need specification to predict scaled solutions against $T_n$.

### 5.3. Imbibition behavior from normalized CDCs

Saturation functions from Kleppe and Morse (1974) were adapted to correlations (35) to (36), **Figure 3**. Function- and system parameters are listed in **Table 1** (tables are at the end of the paper). Assuming five oil viscosities $\mu_o$ (0.01 to 100 cP) gives CDCs $D(S_n)$ and after normalization $\Lambda_n(S_n)$, **Figure 3**. Increasing $\mu_o$ four orders of magnitude reduces $\bar{D}$ and $\tau$ increases accordingly, by a factor 12 (**Table 2**). The peak shifts to lower $S_n$.

Considering $\Lambda_n$ (same mean of 1), only the shape is shifted, from having most of the coefficient collected around a high saturation peak (the 0.01 cP case) to more even distributions at high viscosity. The change is quantified by lower $z_{0,1}$ and $z_{0.5,1}$ ($z_{0,0.5}$ did not change much) as $\Lambda_n$ shifts to lower saturations (**Table 2**).

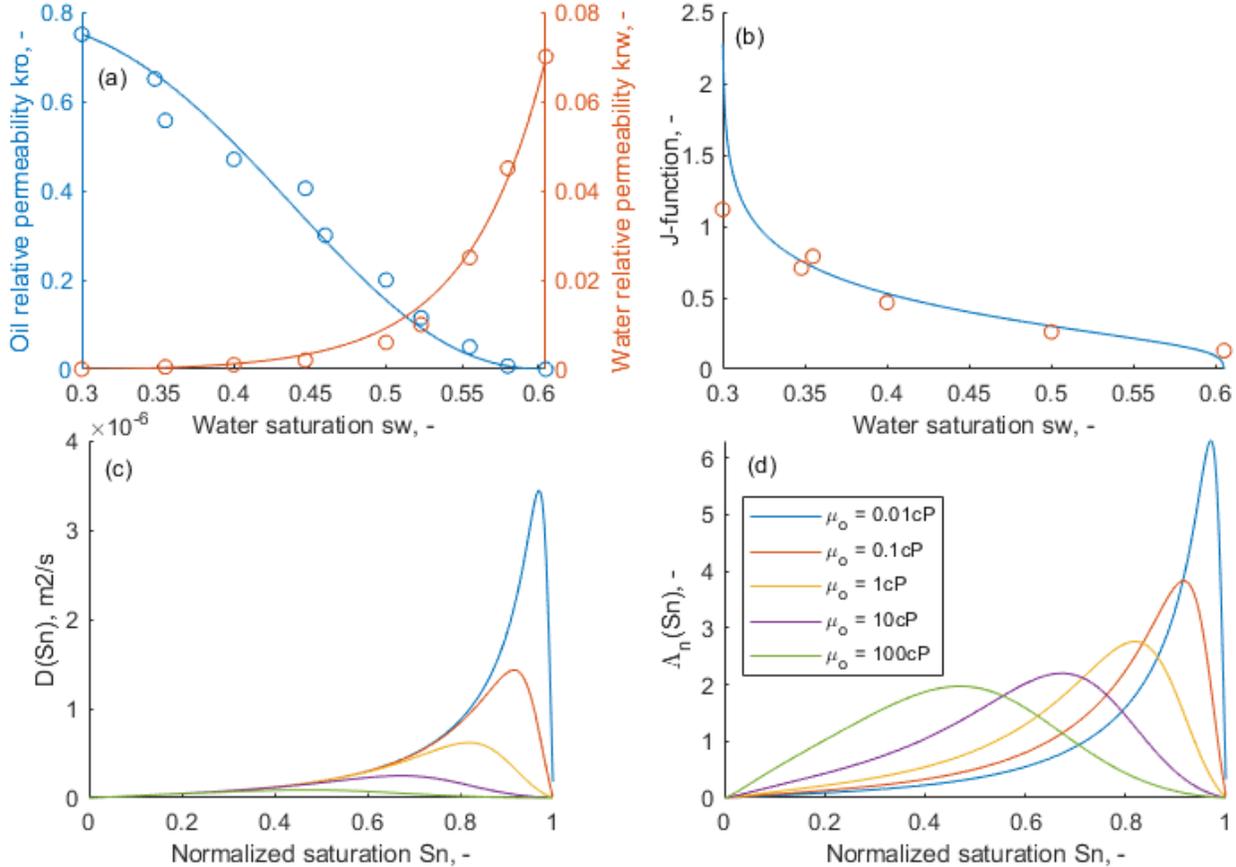

**Figure 3** Input relative permeabilities (a) and J-function (b) based on Kleppe and Morse (1974) (points), adapted to correlations (lines) (35) to (36). Corresponding CDCs $D(S_n)$ in (c) and normalized CDCs $\Lambda_n(S_n)$ in (d) for different oil viscosities.

Numerical solutions of $RF(T^{0.5})$ are shown in **Figure 4a** based on each $\Lambda_n(S_n)$. In all cases $RF$ is linear with $T^{0.5}$ initially and then falls below the extended straight line. For $\Lambda_n(S_n)$ with higher $z_{0,1}$ (lower oil viscosity) the early time recovery goes faster (higher slope $2A$ and equivalently lower $T_{ch}$) and lasts longer in terms of both $RF_{cr}$ and $RF_{tr}$ (marked with circles and triangles respectively), quantified in **Table 2**.

$RF_{tr}$ is clearly higher than $RF_{cr}$ (by ~0.15 units) and visually better represents the deviation from straight line behavior. In **Figure 4b** we plot $RF$ against $T_n^{0.5}$. All curves fall on the same line $RF = T_n^{0.5}$ at early time but deviate from the line at different $RF_{tr}$.

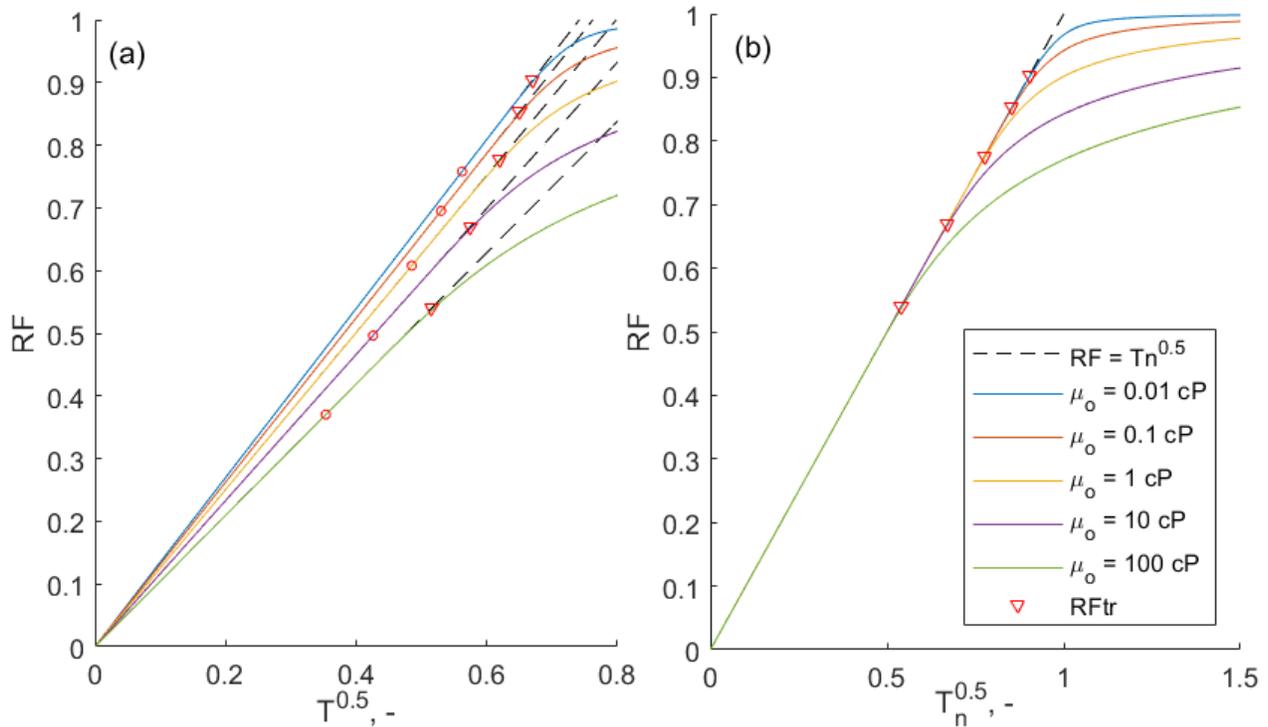

**Figure 4** Recovery RF against $T^{0.5}$ (a) and $T_n^{0.5}$ (b) based on $\Lambda_n$ with five choices of oil viscosity (0.01 cP to 100 cP). $RF_{cr}$ (circles), and $RF_{tr}$ (triangles) are marked. Dashed lines indicate extended early-time solutions (equivalent to $RF = T_n^{0.5}$).

In **Figure 5**, $RF$ is plotted against time (hours) together with the correlation proposed to describe recovery. The lowest $R^2$ of the five cases was 0.9999 and the highest $RMSE$ was 0.0021. The applied characterization thus appears to describe recovery well. The 10 cP case in **Figure 5** was also simulated using IORCoreSim (Lohne 2013), for validation of the numerical code.

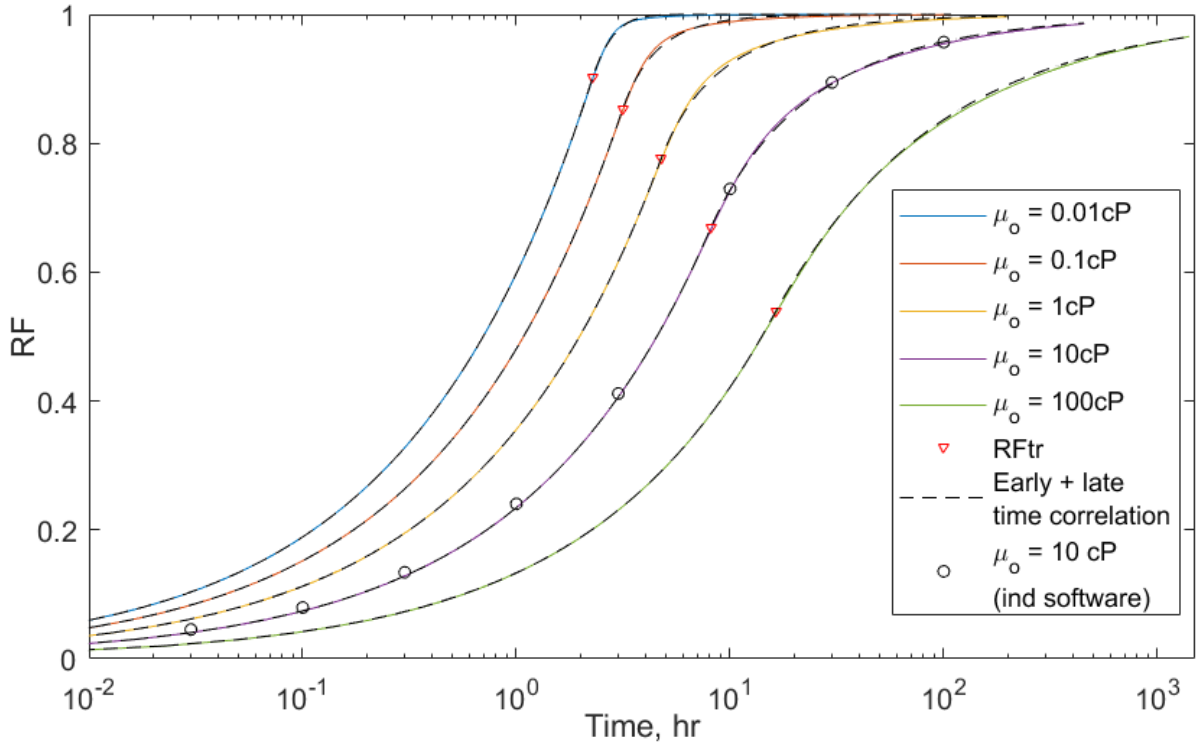

**Figure 5 Numerically calculated (full lines) RF against time with corresponding correlations (dashed lines) and $RF_{tr}$ indicated (red triangles). The 10 cP case was also simulated with an independent software (circles).**

Numerically calculated saturation profiles are shown in **Figure 6**. Estimated profiles based on (32) are also provided, at and before critical time. At low oil viscosity 0.01 cP (high $z_{0,1}$) saturations are high behind the front (given by a large $m = 3.1$), while at high oil viscosity 100 cP (low $z_{0,1}$) the saturations fall quickly near the inlet (reflected by a low $m = 0.588$). The estimated profiles follow the numerical solution profiles reasonably and capture that at higher $z_{0,1}$ a higher recovery is obtained when the no-flow boundary is reached, or equivalently that at high $z_{0,1}$ the fastest saturation $S_n = 0$ is not very fast compared to the remaining profile. Note that our main focus is on recovery (which is well predicted), and not the saturation profiles (where better prediction is possible).

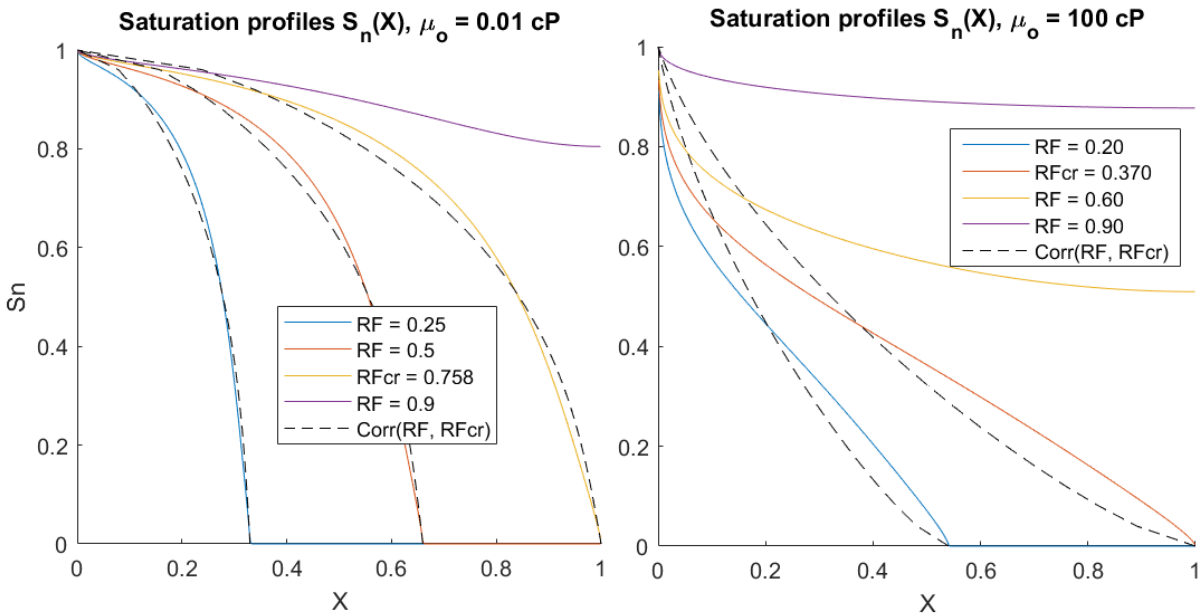

Figure 6 Profiles $S_n(X)$ at different levels of $RF$ for $\mu_o = 0.01$ cP (high $z_{0,1}$) (a) and 100 cP (low $z_{0,1}$) (b). Full lines are calculated numerically, while dashed lines are shown at or before critical time based on (32).

### 5.4. Simulation database

We generate 5500 functions $\Lambda_n(S_n)$ from 5500 random combinations of the seven parameters $n_{w1}, n_{w2}, n_{o1}, n_{o2}, S_{eq}, \log_{10}\left(\frac{J_1}{J_2}\right), \log_{10}\left(\frac{k_{ro}^* \mu_w}{k_{rw}^* \mu_o}\right)$, see **Table 3**. Half the cases were SWW by setting $S_{eq} = 0.999$, while the others had random values down to $S_{eq} = 0.2$. $\log_{10}\left(\frac{J_1}{J_2}\right)$ was positively correlated with $S_{eq}$ since the positive $J$-term should be more dominant in more water-wet systems.

Based on each $\Lambda_n(S_n)$ we solved the model numerically and semi-analytically to calculate $RF$ at early and late scaled times $T$ and $T_n$ and quantified $z_{0,1}, z_{0,0.5}, z_{0.5,1}$ (for the coefficient shape) and $A, RF_{cr}, RF_{tr}, lr$ (characterizing recovery). Each curve $RF(T_n)$ described by the correlation (26) and (30) with $RF_{tr}$ and $lr$ was compared with the curve (numerical solution) it was approximating. The mean $RMSE$ was 0.0045 and the mean $R^2$ was 0.9989. The histograms in **Figure 7** indicate that $RMSE < 0.01$ for 90% of the cases and $R^2 > 0.995$ for 95% of the cases and good match also on the outliers. We thus find the correlation to be an accurate representation of COUSI recovery and that understanding what affects the related parameters is key to describing what affects COUSI behavior in general.

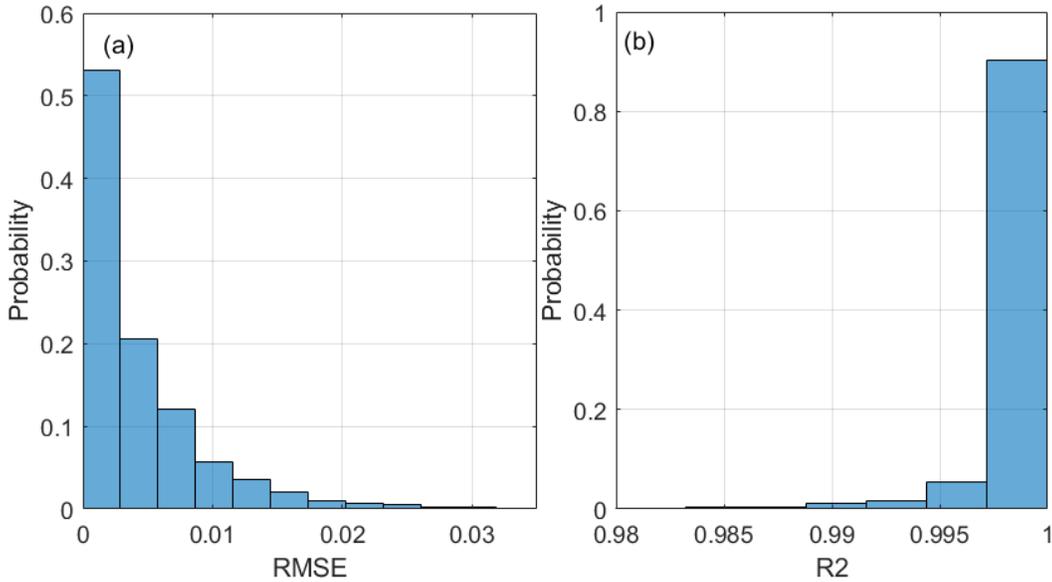

Figure 7 Quantitative description of how well the correlation (26) and (30) matches numerically calculated recovery profiles for the 5500 simulations.

### 5.5. Parameter correlations

Nonlinear regression correlations referred to in the following were obtained using multivariable polynomials. Expressions, metrics and cross plots are found in **Supp Mat Section D**.

#### 5.5.1. Estimation of imbibition parameters

Based on the 5500 simulations, $A, RF_{tr}$ and $RF_{cr}$ were plotted against $z_{0,1}$ in **Figure 8**. They all predominantly increase with $z_{0,1}$, in line with **Section 5.3**, but also show scatter for a fixed $z_{0,1}$. The data could be sorted vertically (at fixed $z_{0,1}$) by $z_{0,0.5}$ when $z_{0,1} < 0.35$ for $A$ and $RF_{tr}$ and $z_{0,1} < 0.85$ for $RF_{cr}$ and by the value of $z_{0.5,1}$ when $z_{0,1} > 0.35$ for $A$ and $RF_{tr}$ and $z_{0,1} > 0.85$ for $RF_{cr}$.

These trends directly state that higher $RF$ is obtained in the square root regime, whether defined by critical time recovery (with $RF_{cr}$) or transition time recovery ($RF_{tr}$), when $\Lambda_n$ is shifted to higher saturations (quantified by increased $z_{0,1}$ or increase in the second fraction at a fixed $z_{0,1}$). $RF_{tr}$ was consistently higher than $RF_{cr}$, by 0.05 to 0.2 units for 90% of the cases (**Figure 8d**), indicative that the transition period after critical time is significant.

The early time imbibition rate coefficient $A$, similarly increases as $\Lambda_n$ is shifted to higher saturations. $A$ spanned a narrow range from 0.2 to 0.7. Thus, $\Lambda_n$ determines $A$ within a factor 3.5, and $T_{ch} = 1/4A^2$ within a factor of 12. $A$ only contains the contribution from the *shape* of $D$ on time scale. The remaining contribution is from $\tau = L^2/\bar{D}$.

Correlations were developed for $A$, $RF_{tr}$ and $RF_{cr}$ as function of $(z_{0,1}, z_{0,0.5})$ or $(z_{0,1}, z_{0.5,1})$ on the stated ranges of $z_{0,1}$, with $R^2$ varying from 0.989 to 0.998.

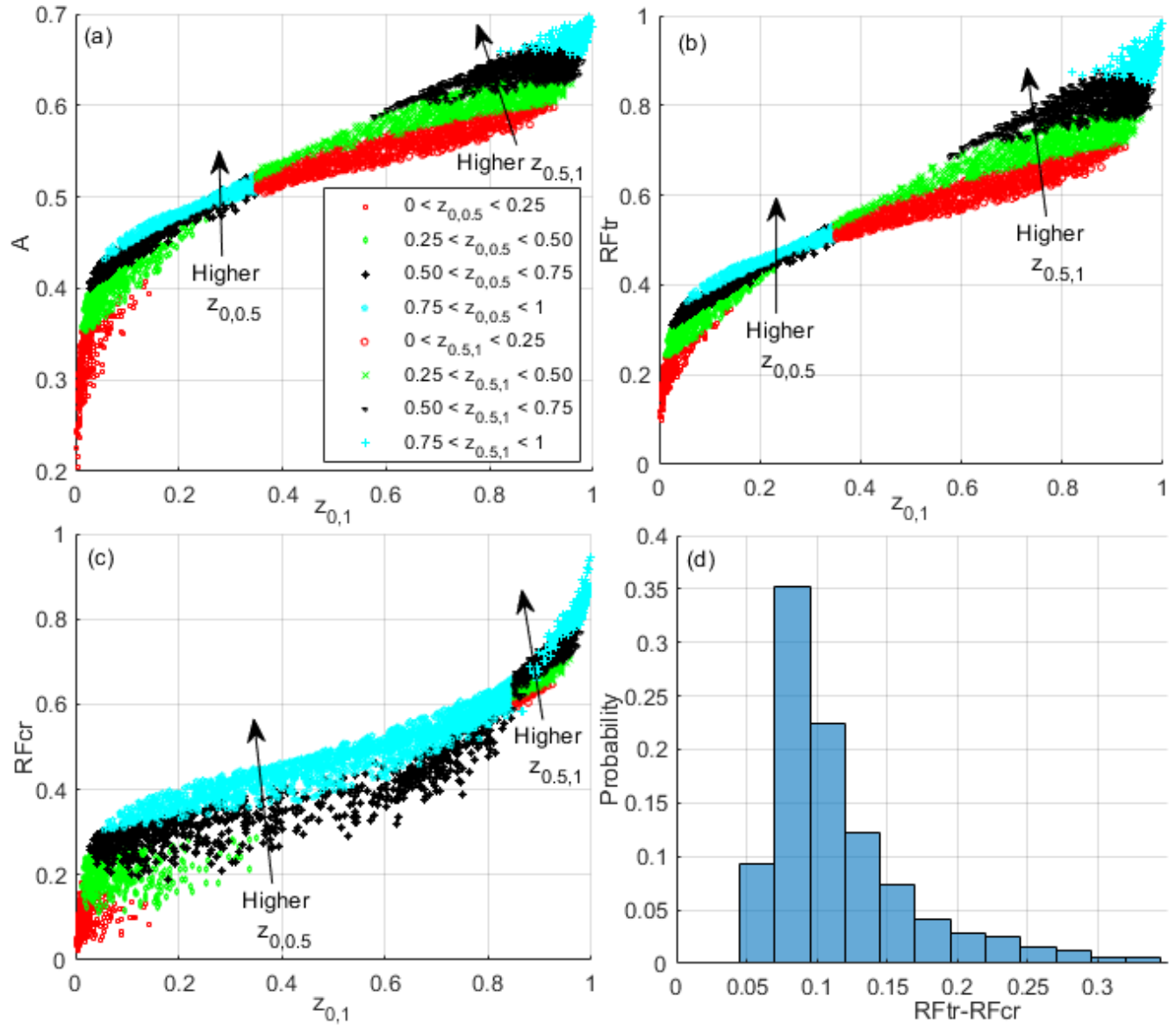

**Figure 8** $A$ (a), $RF_{tr}$ (b) and $RF_{cr}$ (c) plotted against $z_{0,1}$ and sorted by their values of $z_{0,0.5}$ for low $z_{0,1}$ and their values of $z_{0.5,1}$ for high $z_{0,1}$. In (d) a histogram of the differences $RF_{tr} - RF_{cr}$.

The parameter $lr$ is plotted in **Figure 9a** against $z_{0.5,1}$ and sorted into water-wet cases (WW) in blue and mixed-wet (MW) cases in red defined by whether $S_{eq} > 0.99$ or not, respectively. This sorting was selected and found useful since it appears late time behavior is controlled by the diffusion coefficient at the highest saturations, which is described particularly by $z_{0.5,1}$ and wetting state. The MW values at a given $z_{0.5,1}$ are significantly higher than the WW values, and for both cases $lr$ increases with $z_{0.5,1}$.

Both these trends relate to fluid mobility at the highest saturations $S_n \approx 1$ (giving larger values of the coefficient): MW cases have $\Lambda_n > 0$ at $S_n = 1$, while WW cases have a zero value (since the oil mobility terminates). Higher $z_{0.5,1}$ indicates how much of $\Lambda_n$ at high saturations is located at the very highest saturations. Thus, if there is high fluid mobility (giving high $\Lambda_n$) at the highest saturations, $lr$ is high and $RF$ quickly approaches 1 at late time. On the other hand, low mobility at the highest saturations gives low $lr$ and $RF$ more slowly approaches 1.

For an illustration of this phenomena, consider the CDCs $\Lambda_n$ in **Figure 3d**. The 100 cP oil case is almost flat at high saturations and $lr$ takes the low value of $-0.23$ (**Table 2**), while as viscosity reduces $\Lambda_n$ is shifted to higher saturations, and $lr$ changes accordingly reaching 1.5 at the lowest viscosity. **Figure 5** demonstrates that this difference in $lr$ results in several orders of magnitude longer time in the late time regime for the high oil viscosity case.

In order to build predictive correlations estimating $lr$, the MW and WW data were treated separately and plotted in **Figure 9b** and **c** as $RF_{tr}$ and $z_{0,1}$ points taking specific values of $lr$. $lr$ varied systematically with $z_{0,1}$ for a given $RF_{tr}$. That was used as a criterion (dashed lines) for whether $lr > 1.5$ (blue points in **Figure 9b** and **c**) which would result in an exponential recovery profile. 98.5 % of the MW points with $q = z_{0,1} - 2.1 RF_{tr} < -0.85$ and 89.2 % of the WW points with $q < -0.98$ had $lr > 1.5$.

$lr$ was well predicted at $RF_{tr} < 0.6$ with $RMSE \sim 0.03$, while at high $RF_{tr} > 0.6$ the prediction was less accurate with $RMSE \sim 0.2$. Accurate $lr$ prediction at low $RF_{tr}$ is however considered more important since a greater portion of recovery then is in the late time regime which is determined by $lr$.

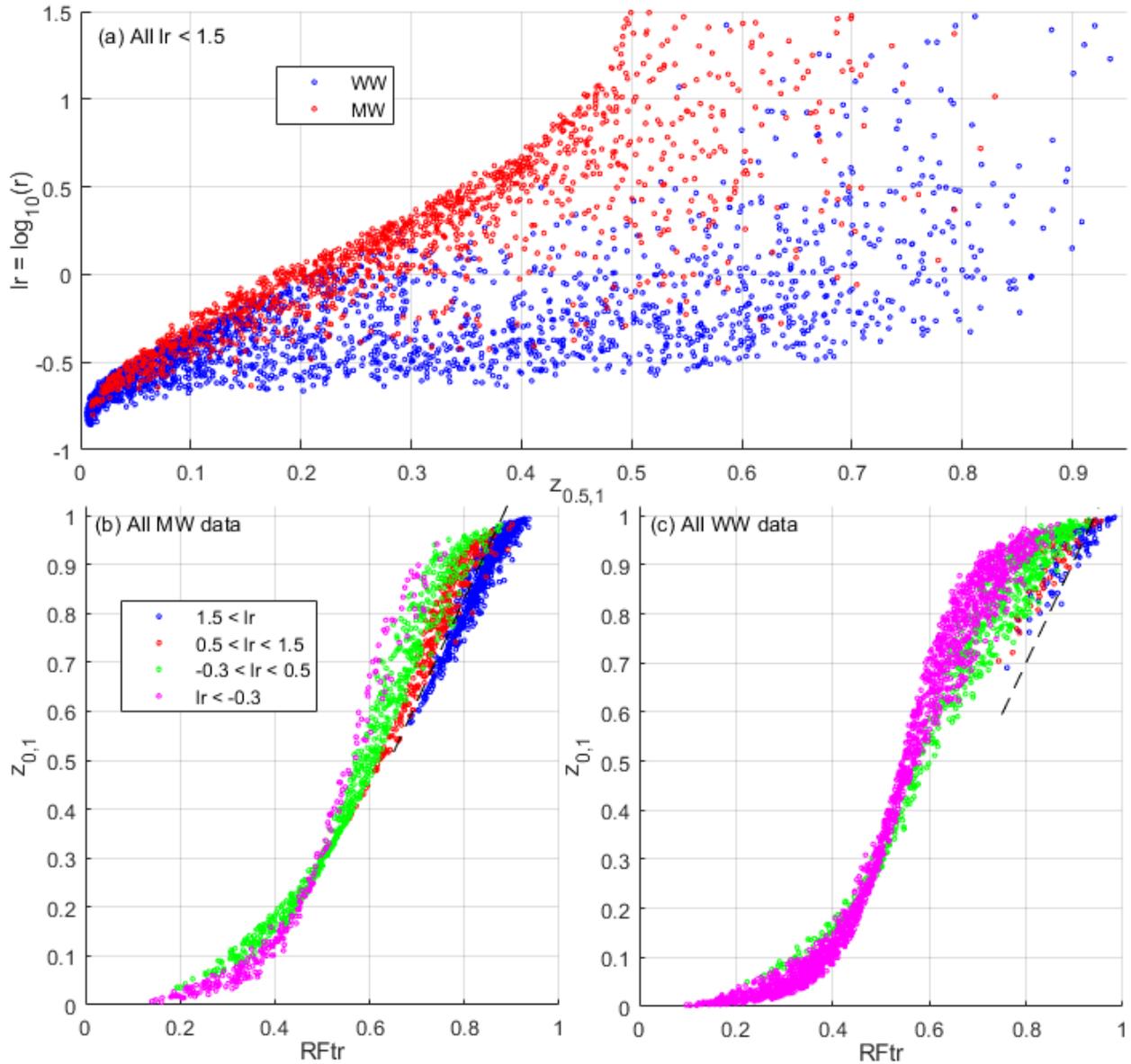

**Figure 9** The $lr$ data plotted against $RF_{tr}$ sorted into MW and WW (a). Maps of $z_{0,1}$ and $RF_{tr}$ with specific value ranges of $lr$ for MW (b) and WW (c) data. Dashed lines indicate the transition to points with $lr > 1.5$.

### *5.5.2. Estimation of diffusion coefficient parameters from recovery parameters*

The parameters $A, z_{0,1}, z_{0.5,1}, z_{0,05}$ are plotted against $RF_{tr}$ in **Figure 10**. $A$ did not display significant scatter and could be estimated accurately as a correlation of $RF_{tr}$ (with $R^2 = 0.9996$). The fractions $z_{a,b}$ generally increase with $RF_{tr}$, predicting a coefficient shifted to higher saturations. They show scatter for a given $RF_{tr}$ but were relatively well sorted by $lr$. $z_{0,1}$ and $z_{0.5,1}$ appeared well defined given $RF_{tr}$ and $lr$. $z_{0.5,1}$ depended mainly on $RF_{tr}$ at $RF_{tr} > 0.8$ and mainly on $lr$ at $RF_{tr} < 0.6$. For all three parameters, but especially $z_{0,05}$, the data were better sorted by $lr$ at $RF_{tr} < 0.6$ than at $RF_{tr} > 0.6$. Correlations for the $z_{a,b}$ fractions were developed as function of $(RF_{tr}, lr)$ on the two intervals (data values of $lr > 1.5$ were set as 1.5 since the recovery curve becomes the same).

The prediction performance was good for $z_{0,1}$ and $z_{0.5,1}$ ($RMSE \sim 0.01$ for $RF_{tr} < 0.6$ and $\sim 0.03$ for $RF_{tr} > 0.6$ for both fractions) while $z_{0,05}$ was estimated less accurately ($RMSE \sim 0.06$ for $RF_{tr} < 0.6$ and $RMSE \sim 0.07$ for $RF_{tr} > 0.6$). Based on recovery profile parameters we are thus able to determine

CDC parameters. That is demonstrated in **Section 5.7** using experimental data.

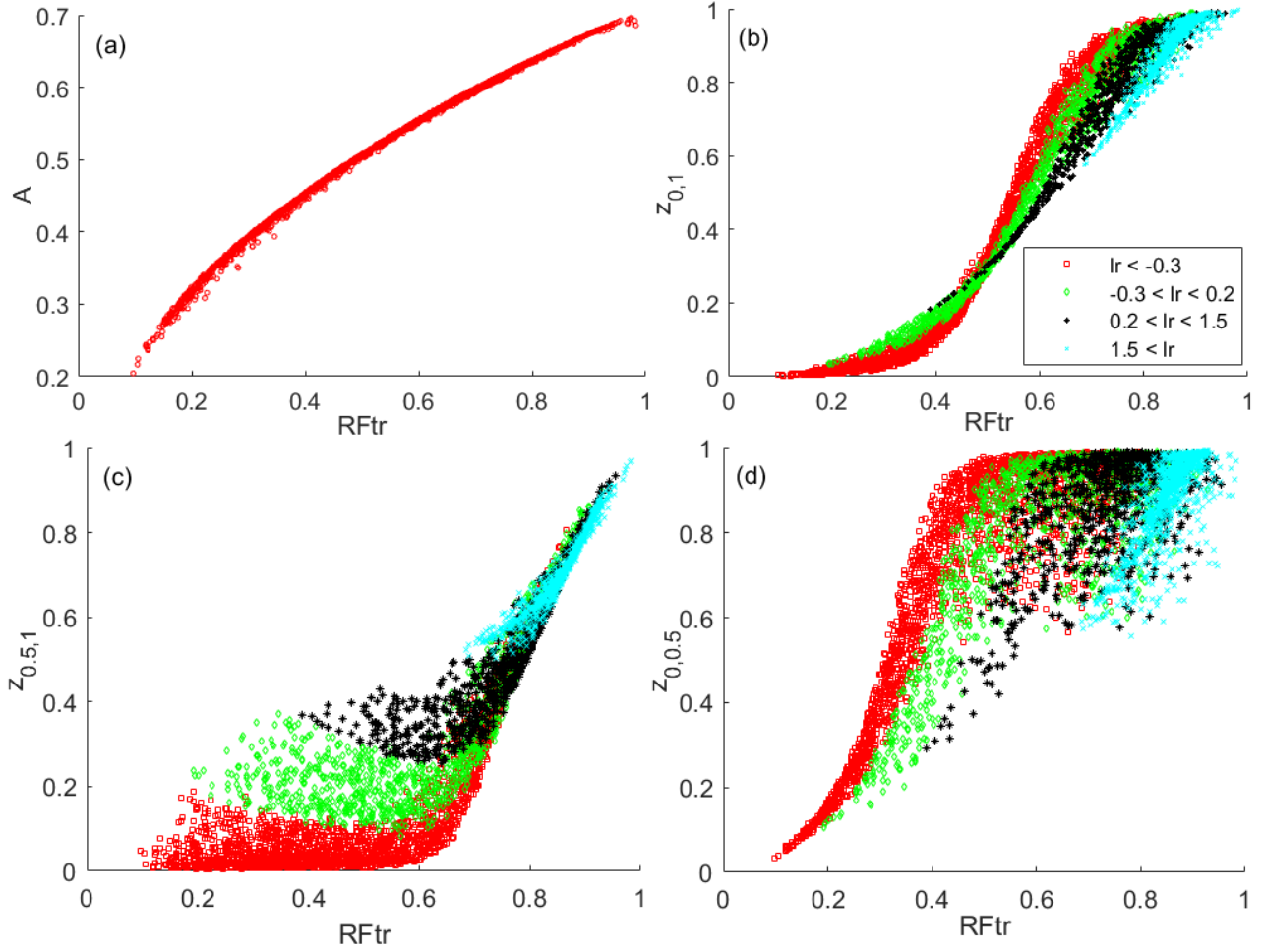

**Figure 10 Plot of $A$ (a), $z_{0,1}$ (b), $z_{0.5,1}$ (c) and $z_{0,05}$ (d) against $RF_{tr}$, the latter three sorted by $lr$.**

## 5.6. Estimation of recovery from the capillary diffusion coefficient

We illustrate estimation of recovery based on diffusion coefficients $D$. Consider the coefficients in **Figure 3c**, based on Kleppe and Morse (1974). They were scaled by $\bar{D}$ to obtain $\Lambda_n$. The three fractions $z_{a,b}$ for each were used to estimate $A, RF_{tr}, lr$ (correlations in **Supp Mat Section D**) giving $RF(T_n)$. To calculate $RF$ against time $t$ we use that $t = \tau T_{ch} T_n$ where $\tau = L^2/\bar{D}$ and $T_{ch} = 1/4A^2$. The estimated curves are plotted against numerically simulated results in **Figure 11a**.

Additionally, we consider MW relative permeabilities and $J$-function from Behbahani and Blunt (2005) (**Figure 12a**). They ran pore scale simulation of wetting conditions and matched experiments by Zhou et al. (2000) with upscaled functions. We vary oil viscosity (from 0.1 to 1000 cP), obtain $\Lambda_n$ (**Figure 12b**), calculate $z_{a,b}$, estimate $A, RF_{tr}$ and $lr$ (**Table 4** and **Table 5**), unscale $T_n$ to get $RF(t)$ and compare with numerical simulations in **Figure 11b** (the two cases with lowest viscosity are similar).

For both datasets $RF(t)$ is well predicted: $RMSE$ and $R^2$ spanned 0.001-0.009 and 0.999-1.000 for the WW example and 0.001-0.005 and 0.999-1.000 for the MW example, respectively.

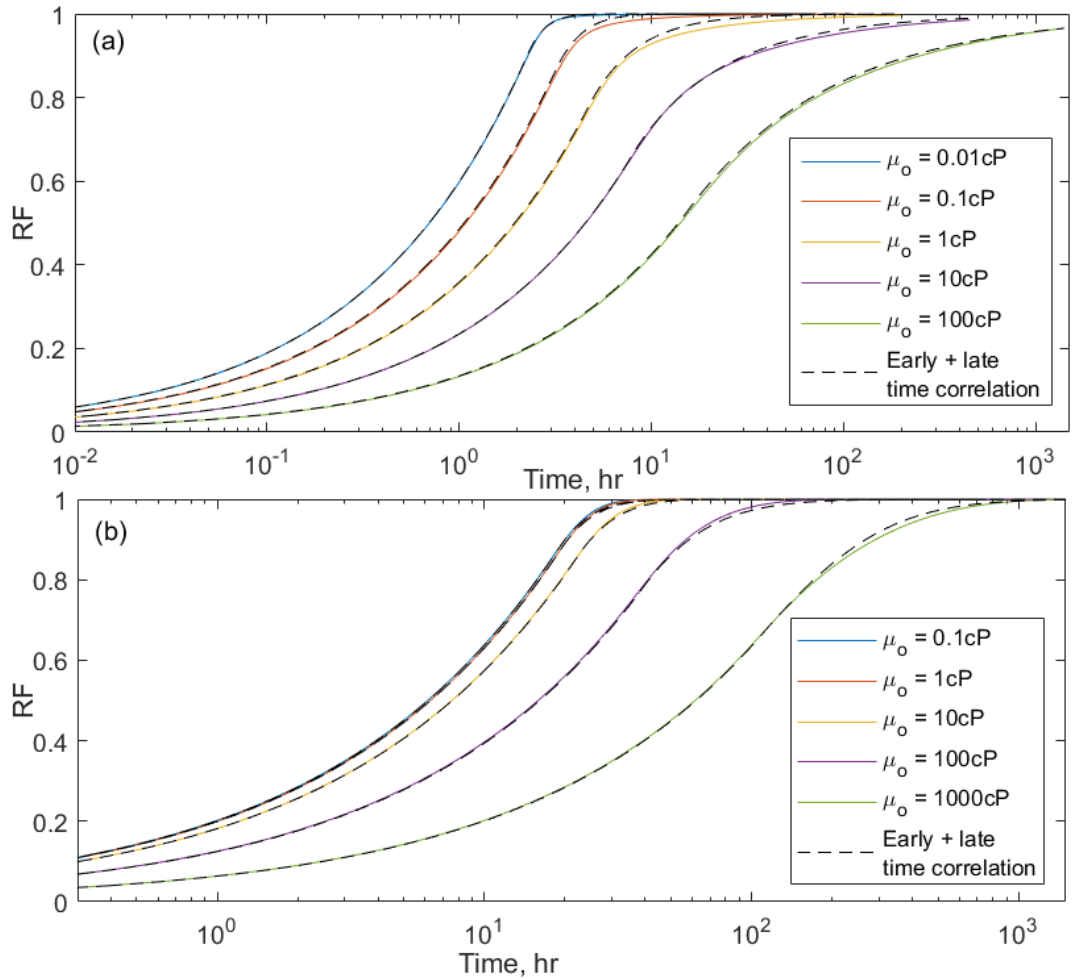

**Figure 11** Predicted $RF(t)$ using correlations (dashed lines) with $A, RF_{tr}, lr$ based on diffusion coefficient fractions $z_{a,b}$ and numerically calculated $RF(t)$ (full lines). Five cases are shown with different oil viscosity and input from Kleppe and Morse (1974) (a) and Behbahani and Blunt (2005) (b).

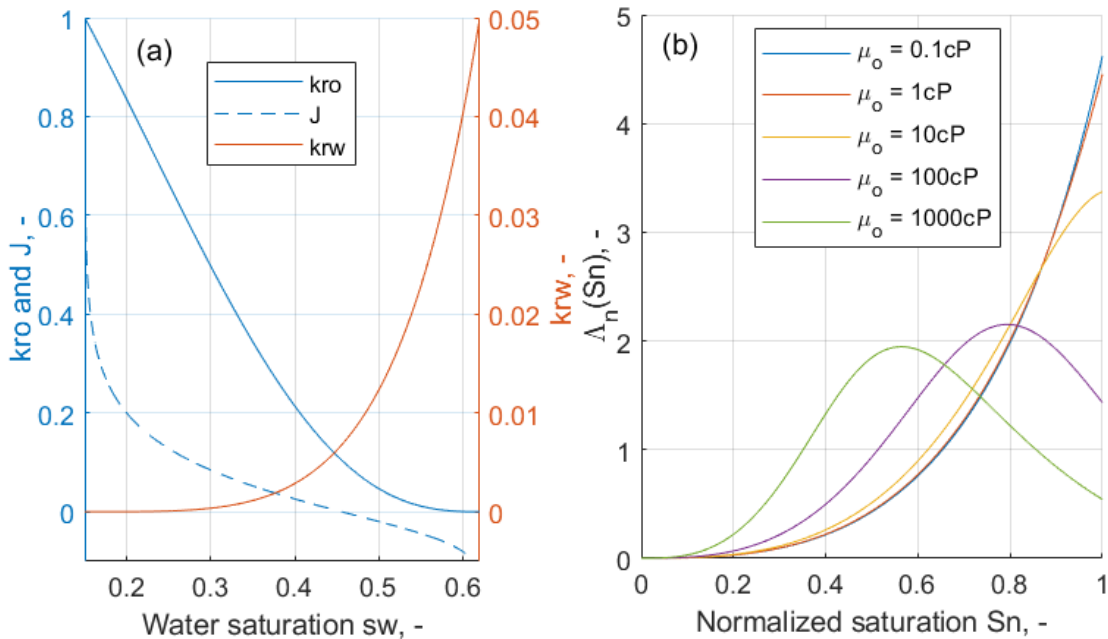

**Figure 12** Saturation functions from Behbahani and Blunt (2005) (a) and resulting $\Lambda_n(S_n)$ for different oil viscosities (b).

## 5.7. Estimation of capillary diffusion coefficients from recovery data

Assume data $RF(t)$ for an imbibition experiment. The data can be plotted against square root of normalized time such that early time data lay on the straight line $RF = T_n^{0.5}$. Based on the plot of $RF$ against $T_n^{0.5}$, we determine $RF_{tr}$ where $RF$ deviates from the straight line. Fitting $RF > RF_{tr}$ (late time data) to (30) provides the value of $lr$. The factor between time $t$ and $T_n$ from scaling equals the product of $\tau$ and $T_{ch}$, which are both unknown. $T_{ch}$ is estimated from the correlation $A(RF_{tr})$ and using $T_{ch} = 1/4A^2$. After, $\tau$ can be calculated. The only unknown parameter in $\tau$, $\bar{\Lambda}$, is calculated subsequently. The fractions $z_{0,1}, z_{0,0.5}, z_{0.5,1}$ characterizing $\Lambda_n$ are estimated from $RF_{tr}$ and $lr$. We then have information about the shape and magnitude of $\Lambda_n, \Lambda$ and $D$.

### 5.7.1. Interpretation of experimental data

Fischer et al. (2008) conducted one-end-open COUSI experiments on SWW Berea sandstone core plugs. $\mu_o$ was either 3.9 or 63.3 cP and wetting phase viscosity was varied from 1.0 to 494.6 cP in five tests each. Recovery was scaled by the highest observed value (48.7%) and plotted against square root of $T_n = \frac{t}{\tau T_{ch}}$ in **Figure 13** by selecting values of $\tau T_{ch}$ for each test.

The early data overlapped as $RF = T_n^{0.5}$. For the 3.9 cP oil tests all the data appear linear, with abrupt deviations at $RF \geq 0.95$, which could be from variation in residual saturation. The high $RF_{tr}$ is consistent with relatively low or similar oil viscosity compared to water viscosity. Water-wet systems tend to have much lower $k_{rw}$ than $k_{ro}$ (Mungan 1972; Kleppe and Morse 1974; Anderson 1987b; Bourbiaux and Kalaydjian 1990; Andersen et al. 2022) so the oil-to-water mobility ratio is high, shifting $\Lambda_n$ to high saturations and $RF_{tr}$ to high values. In the 63.3 cP oil tests, the oil-to-water mobility ratios are lower, and the curves deviate from the linear trend between $RF_{tr} = 0.75$ and 0.9. Only a few tests have significant amounts of late time data. For some such tests we estimate CDCs to explain the observed recovery.

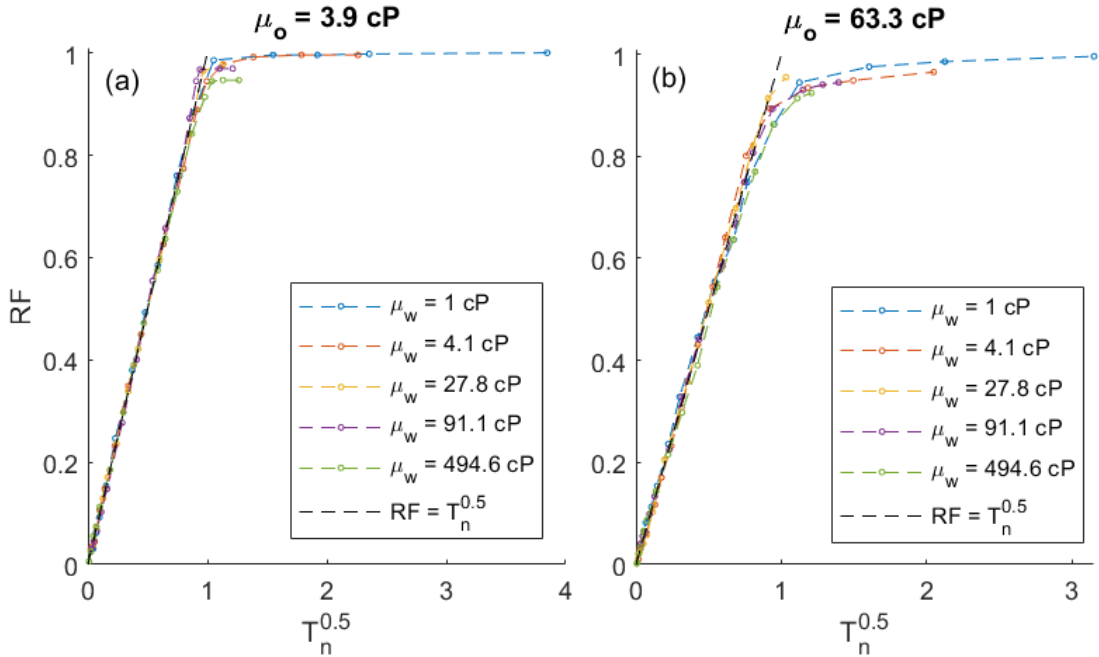

**Figure 13** Experimental data from Fischer et al. (2008) plotted against $T_n^{0.5}$, with varied wetting phase viscosity for oil viscosity of 3.9 cP (a) and 63.3 cP (b).

Three tests with $\mu_o = 63.3$ cP and $\mu_w = 1, 4.1$ and 27.8 cP were matched with $RF_{tr}$ and $lr$, see **Figure 14**. Straight line behavior, deviation from the straight line and late time trends are well captured. The factor $\tau T_{ch}$ was found from time normalization. $A$, and thus $T_{ch}$, were estimated from $RF_{tr}$. Knowing the constants in $\tau$ and $T_{ch}$ we calculate $\bar{\Lambda}$ from $\tau T_{ch}$. Fractions $z_{a,b}$ are calculated from $RF_{tr}$ and $lr$. See matched and estimated parameters in **Table 6**.

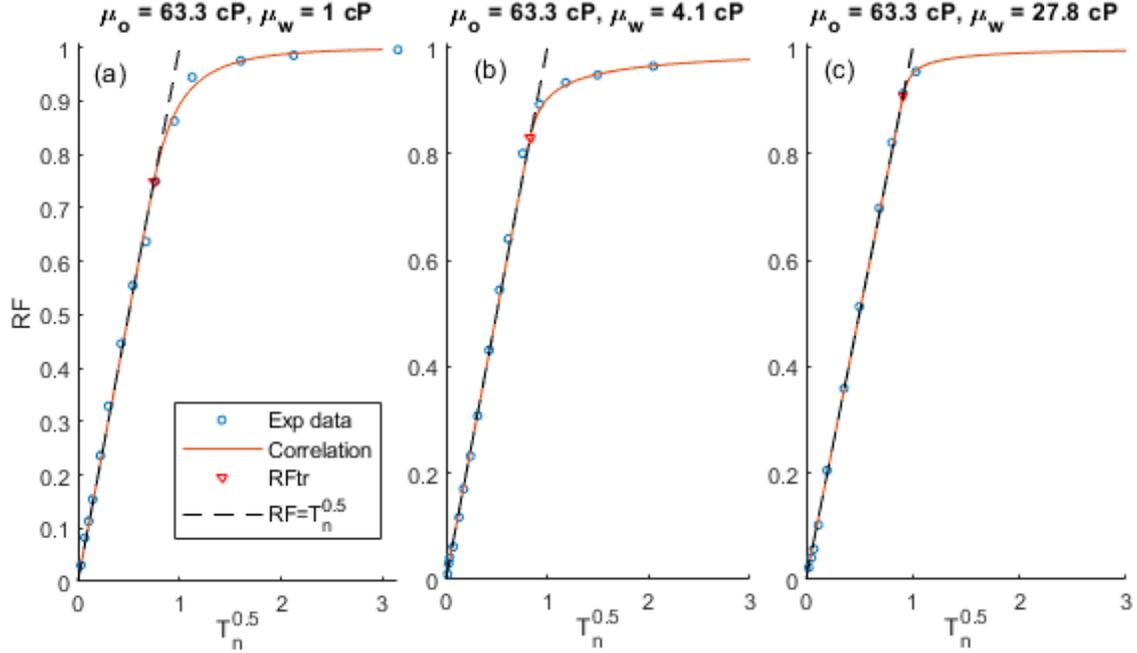

**Figure 14 Experimental RF (points) from Fischer et al. (2008) with same oil viscosity and different water viscosities plotted against $T_n^{0.5}$ and matched to the recovery correlation.**

A function $\Lambda_n$ fitting the fractions $z_{a,b}$ of an experiment can be determined tuning (37) freely (we set $S_{eq} = 0.999$). However, as the three experiments were performed under same conditions (apart from $\mu_w$) we assume wettability, and hence saturation functions are fixed. We thus require fixed parameters $n_{w1}, n_{w2}, n_{o1}, n_{o2}, \frac{J_1}{J_2}, \frac{k_{ro}^*}{k_{rw}^*}, J_2 k_{ro}^*$ when matching the three tests. The last product appears in $\Lambda(S_n)$ and controls $\bar{\Lambda}$. Initial water saturation was zero such that $k_{ro}^* = 1$. We could thus determine $J_1, J_2, k_{rw}^*$ separately, **Table 7**. The mean *relative* error of the matched fractions was 2 to 12 % for the three experiments, i.e. the functions were well adapted (better match is possible by tuning each experiment freely). The estimated $\Lambda_n$ are shown in **Figure 15**. Under the constraint of fixed saturation functions, higher $\mu_w$ shifts $\Lambda_n$ to higher saturations and $\bar{\Lambda}$ to lower values.

For validation, $RF(T)$ is calculated solving the PDE with the determined $\Lambda_n$. $\tau$ uses $\bar{\Lambda}$ corresponding to the determined saturation functions and viscosities, $A$ follows analytically from $\Lambda_n$. $RF$ against $t$ and $T_n^{0.5}$ compared with experimental data is shown in **Figure 16**. The forward simulations overlap the experimental data very well, but late time estimated $RF$ is higher than experimental points for the 4.1 cP test. Higher water viscosity consistently leads to higher $RF_{tr}$ when simulated, but the 4.1 cP test did not follow this experimentally (like the other two tests), which is attributed to experimental variation (in e.g. residual saturation). We should expect monotonous trends when the same input are used except only the oil viscosity is varied.

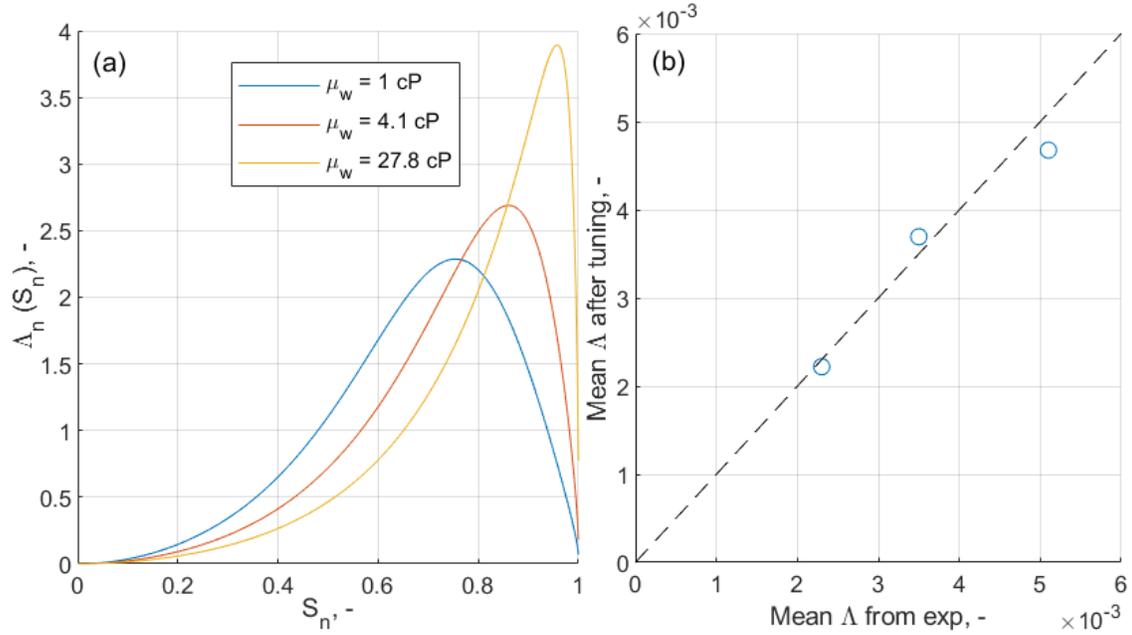

**Figure 15** Estimated $\Lambda_n(S_n)$ (a) and comparison of estimated vs experimental mean of coefficients $\Lambda(S_n)$ (b). The coefficients are based on estimated fractions $z_{a,b}$ and time scaling from Fischer et al. (2008)'s experiments, constrained by assuming same saturation functions.

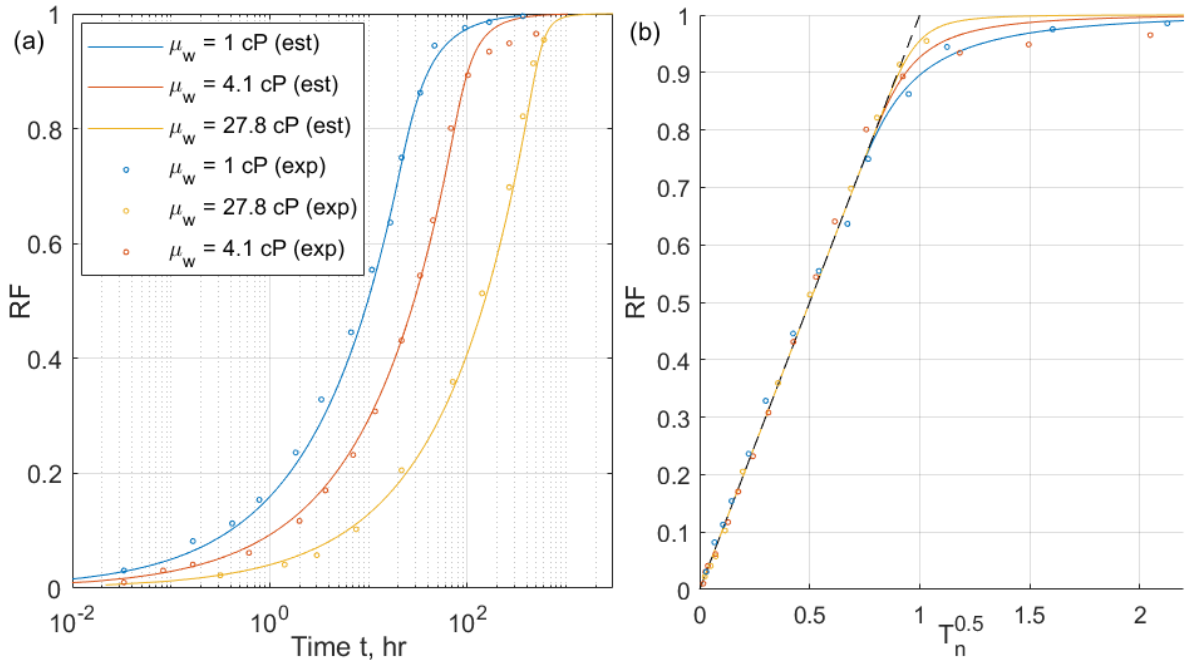

**Figure 16** Comparison of RF from forward simulation of estimated coefficients $\Lambda_n$ and RF data from Fischer et al. 2008, plotted against log time (a) and against $T_n^{0.5}$ (b).

## 6. Conclusions

The 1D linear COUSI problem was normalized and shown to depend only on a normalized diffusion coefficient $\Lambda_n(S_n)$. By investigating how variations of this coefficient impacts recovery we practically determine what controls COUSI. Based on theory and running 5500 cases spanning expected possible shapes of $\Lambda_n(S_n)$ and investigating the resulting recovery curves, the following findings were made:

- Scaled recovery could be described as equal to the square root of scaled time $RF = \sqrt{T_n}$ in an early time regime until a limit $RF_{tr}$; and by a late time regime described by an Arps type equation with a parameter $lr$ controlling how fast the imbibition rate declines. Our definition of transition from early to late time seems to explain the good match (mean $R^2 = 0.9989$) compared to previous works using similar descriptions.
- $RF_{tr}$ (the recovery when deviation is seen from square root of time trend) is significantly higher than $RF_{cr}$ (the recovery when water reaches the no-flow boundary). The difference was mainly between 0.05 and 0.2 units.
- $T_n$ is a universal scaled time for COUSI. It relates to regular time via one factor equal the mean of the diffusion coefficient $D$ divided by squared system length, and a factor $T_{ch}$ depending only on the coefficient shape, i.e. $t = \tau T_{ch} T_n$ and $\tau = L^2/\bar{D}$. $T_{ch}$ varied between 0.5 and 6 meaning $\tau$ alone gives correct time scale for COUSI within approximately one order of magnitude. However, the fact that $RF_{tr}$ and $lr$ take many different values demonstrates that full imbibition profiles do not scale / overlap; only the early-time data.
- It was possible to associate the shape of $\Lambda_n$ with different early- and late time behavior of the recovery profiles and make accurate qualitative and quantitative predictions:
    - Diffusion coefficients $\Lambda_n$ that are shifted to higher saturations have high early time recovery (high $RF_{tr}$) and low $T_{ch}$ (faster imbibition rate due to coefficient shape). $\Lambda_n$ shifts to higher saturations when oil-to-water mobility ratio increases.
    - Diffusion coefficients $\Lambda_n$ with significantly positive values at the highest saturations have more rapid late time recovery (higher $lr$). The mixed-wet data had generally higher $lr$ than the water-wet data since imbibition ends at saturations where both fluids have mobility and thus $\Lambda_n > 0$.
    - Three fractions $z_{a,b}$ were used to describe the shape of $\Lambda_n$. The parameters $T_{ch}, RF_{tr}$ and $lr$ could be predicted from their values and yield full recovery curves matching the solutions obtained by numerical solution of the diffusion equation.
    - These trends, directly related to the shape of $\Lambda_n$ which can be observed visually, give a clear intuitive understanding of what recovery behavior to expect for a given diffusion coefficient, or how recovery will change based on how a parameter changes the diffusion coefficient.
- Characteristic features of recovery profiles could also be used to quantify the capillary diffusion function needed to get such recovery:
    - Recovery data can be fitted to be described by $RF_{tr}$ and $lr$, which identify and quantify the early and late time regimes. A high $RF_{tr}$ indicates that $\Lambda_n$ should be shifted to high saturations, while a high $lr$ indicates how significantly $\Lambda_n$ is shifted towards the highest saturations.

## Statements & Declarations

**Funding**: The author declares that no funds, grants, or other support were received during the preparation of this manuscript.
**Competing Interests**: The author has no relevant financial or non-financial interests to disclose.
**Author Contributions**: P. Ø. Andersen has made all contributions in this paper.
**Data Availability**: The datasets generated or analyzed in the current study are available in the cited references or can be obtained by following the outlined instructions of the paper.

## Tables

Table 1 Input parameters characterizing the system and saturation functions from Kleppe and Morse (1974).

| | | | | | |
|---|---|---|---|---|---|
| $n_{w1}$ | 6 | $J_1$ | 0.3 | $K$ | 290 mD |
| $n_{w2}$ | 2.5 | $J_2$ | 0.03 | $L$ | 0.1 m |
| $n_{o1}$ | 2 | $S_{eq}$ | 0.999 | $\phi$ | 0.225 |
| $n_{o2}$ | 0.5 | $s_{wr}$ | 0.30 | $\mu_w$ | 1 cP |
| $k_{rw}^*$ | 0.07 | $s_{or}$ | 0.395 | $\mu_o$ | 0.01 cP to 100 cP |
| $k_{ro}^*$ | 0.75 | | | $\sigma_{ow}$ | 21 mN/m |

Table 2 'Imbibition parameters' were calculated from the semi-analytical ($T_{ch}$ and $RF_{cr}$) and numerical ($RF_{tr}$ and $r$) recovery solution. $J_1/J_2 = 10$ while $\frac{k_{ro}^* \mu_w}{k_{rw}^* \mu_o}$ is 1070 at 0.01 cP oil viscosity and reduces to 0.107 at 100 cP. Values without units are dimensionless. 'Matching metrics' describe the correlation fit to the numerical recovery profile.

| Case | Scaling parameters | | | CDC shape parameters | | | Imbibition parameters | | | | Matching metrics | |
|---|---|---|---|---|---|---|---|---|---|---|---|---|
| $\mu_o$ [cP] | $\bar{D}$ [$m^2/s$] | $\bar{\Lambda}$ | $\tau$ [hr] | $z_{0,1}$ | $z_{0.5,1}$ | $z_{0,0.5}$ | $A$ | $T_{ch}$ | $RF_{cr}$ | $RF_{tr}$ | $lr$ | $R^2$ | RMSE |
| 0.01 | 5.46e-7 | 0.70e-3 | 5.1 | 0.934 | 0.822 | 0.796 | 0.674 | 0.550 | 0.758 | 0.903 | 1.5 | 1.0000 | 0.0008 |
| 0.1 | 3.74e-7 | 1.5e-3 | 7.4 | 0.903 | 0.734 | 0.796 | 0.656 | 0.581 | 0.695 | 0.853 | 0.32 | 1.0000 | 0.0016 |
| 1 | 2.24e-7 | 2.9e-3 | 12.4 | 0.839 | 0.552 | 0.796 | 0.627 | 0.637 | 0.608 | 0.776 | 0.03 | 0.9999 | 0.0021 |
| 10 | 1.13e-7 | 4.6e-3 | 24.7 | 0.692 | 0.285 | 0.789 | 0.583 | 0.736 | 0.496 | 0.670 | -0.13 | 0.9999 | 0.0018 |
| 100 | 4.50e-8 | 5.8e-3 | 61.8 | 0.411 | 0.143 | 0.735 | 0.524 | 0.911 | 0.370 | 0.540 | -0.23 | 0.9999 | 0.0016 |

Table 3 Random selection of input parameters to generate normalized CDCs $\Lambda_n(S)$ using uniform distributions with limits indicated. [a] Half the cases were SWW by setting $S_{eq}$ equal to 0.999. [b] $\frac{J_1}{J_2}$ was correlated with $S_{eq}$ such that $\log_{10}\left(\frac{J_1}{J_2}\right) = \frac{S_{eq}-0.2}{0.8} + rand(-1,1)$.

| | min | max | | min | max |
|---|---|---|---|---|---|
| $n_{w1}$ | 0.5 | 6 | [a] $S_{eq}$ | 0.2 | 0.999 |
| $n_{w2}$ | 1.5 | 6 | [b] $\log_{10}\left(\frac{J_1}{J_2}\right)$ | -1 | 2 |
| $n_{o1}$ | 1.5 | 6 | $\log_{10}\left(\frac{k_{ro}^*}{k_{rw}^*}\frac{\mu_w}{\mu_o}\right)$ | -3.5 | 4.5 |
| $n_{o2}$ | 0.5 | 6 | | | |

Table 4 System and saturation function input from Behbahani and Blunt (2005).

| | | | | | |
|---|---|---|---|---|---|
| $n_{w1}$ | 5 | $J_1$ | 0.1 | $K$ | 3131 mD |
| $n_{w2}$ | 4 | $J_2$ | 0.02 | $L$ | 10 cm |
| $n_{o1}$ | 2.5 | $S_{eq}$ | 0.65 | $\phi$ | 0.207 |
| $n_{o2}$ | 1.5 | $s_{wr}$ | 0.15 | $\mu_w$ | 0.967 cP |
| $k_{rw}^*$ | 0.05 | $s_{or}$ | 0.38 | $\mu_o$ | 0.1 to 1000 cP |
| $k_{ro}^*$ | 1 | | | $\sigma_{ow}$ | 0.0242 N/m |

Table 5 Calculated parameters based on MW data from Behbahani and Blunt (2005).

| Case | Scaling parameters | | | CDC shape parameters | | | Estimated imbibition parameters | | |
|---|---|---|---|---|---|---|---|---|---|
| $\mu_o$ [cP] | $\bar{D}$ [$m^2/s$] | $\tau$ [hr] | | $z_{0,1}$ | $z_{0.5,1}$ | $z_{0,0.5}$ | $A$ | $RF_{tr}$ | $lr$ |
| 0.1 | 6.34e-8 | 43.8 | | 0.947 | 0.760 | 0.936 | 0.667 | 0.882 | 1.50 |
| 1 | 6.20e-8 | 44.8 | | 0.946 | 0.760 | 0.936 | 0.667 | 0.880 | 1.50 |
| 10 | 5.24e-8 | 53.0 | | 0.936 | 0.718 | 0.936 | 0.661 | 0.863 | 1.50 |
| 100 | 2.68e-8 | 103.5 | | 0.880 | 0.551 | 0.934 | 0.634 | 0.791 | 0.581 |
| 1000 | 0.805e-8 | 345.2 | | 0.687 | 0.347 | 0.916 | 0.590 | 0.683 | 0.589 |

**Table 6 Parameters matched to experimental recovery data from Fischer et al. (2008) and resulting estimated parameters to determine the CDC and time scales.**

|  | Matched parameters | | | Estimated parameters | | | | | | |
|---|---|---|---|---|---|---|---|---|---|---|
| $(\mu_o, \mu_w)[cP]$ | $RF_{tr}$ | $lr$ | $\tau T_{ch}[h]$ | $A(RF_{tr})$ | $T_{ch}(A)$ | $\tau [h]$ | $\bar{\Lambda}$ | $z_{0,1}$ | $z_{0.5,1}$ | $z_{0,0.5}$ |
| (63.3, 1) | 0.75 | 0.2 | 36.93 | 0.617 | 0.657 | 56.2 | 5.1e-3 | 0.826 | 0.449 | 0.895 |
| (63.3, 4.1) | 0.83 | -0.3 | 119.3 | 0.648 | 0.595 | 201 | 3.5e-3 | 0.949 | 0.688 | 0.956 |
| (63.3, 27.8) | 0.91 | -0.2 | 564.6 | 0.677 | 0.546 | 1034 | 2.3e-3 | 0.967 | 0.907 | 0.995 |

**Table 7 Saturation function parameters used to calculate CDCs with correct fractions and correct $\bar{\Lambda}$ values. Time scales are consistent with calculated saturation functions.**

| Tuned saturation function parameters | | | | | | Matched CDC parameters | | | | Time scale parameters | | |
|---|---|---|---|---|---|---|---|---|---|---|---|---|
| $n_{w1}$ | 5.32 | $k_{rw}^*$ | 0.0026 | $s_{wr}$ | 0.0 | Case | $z_{0,1}$ | $z_{0.5,1}$ | $z_{0,0.5}$ | $\bar{\Lambda}$ | $\tau [hr]$ | $T_{ch}$ | $\tau T_{ch}$ |
| $n_{w2}$ | 3.29 | $k_{ro}^*$ | 1 | $s_{or}$ | 0.513 | 1 cP | 0.831 | 0.462 | 0.889 | 4.7e-3 | 60.5 | 0.651 | 39.4 |
| $n_{o1}$ | 1.54 | $J_1$ | 2.55 | $S_{eq}$ | 0.999 | 4.1 cP | 0.893 | 0.617 | 0.891 | 3.7e-3 | 191.4 | 0.605 | 115.8 |
| $n_{o2}$ | 4.09 | $J_2$ | 0.106 | | | 27.8 cP | 0.931 | 0.745 | 0.891 | 2.2e-3 | 1060 | 0.566 | 600.1 |